\documentclass[journal]{IEEEtran}

\usepackage{cite}
\usepackage{amsmath,amssymb,amsfonts}
\usepackage{algorithmic}
\usepackage{multirow}
\usepackage{array}
\usepackage{graphicx}
\usepackage{textcomp}
\usepackage[caption=false,font=footnotesize]{subfig}
\usepackage{subfig}
\usepackage{stfloats}
\usepackage{url}
\usepackage{xcolor}

\begin{document}

\title{Mitigating Vulnerabilities of\\ Voltage-based Intrusion Detection Systems in\\ Controller Area Networks}

\author{Sang~Uk~Sagong,~\IEEEmembership{Student Member,~IEEE,}
Radha~Poovendran,~\IEEEmembership{Fellow,~IEEE,}
and~Linda~Bushnell,~\IEEEmembership{Fellow,~IEEE}
\thanks{S. Sagong, R. Poovendran, and L. Bushnell are with the Department
of Electrical and Computer Engineering, University of Washington, Seattle, WA, 98195. {\tt \{sagong,rp3,lb2\}@uw.edu}}
\thanks{Part of this work was presented at Embedded Security in Cars Europe in 2018 \cite{Sagong:2018:escar}.}
}

\markboth{}%
{Sagong \MakeLowercase{\textit{et al.}}: Mitigating Vulnerabilities of Voltage-based Intrusion Detection Systems in Controller Area Networks}

\maketitle

\begin{abstract}
	\label{sec:abstract}
	Data for controlling a vehicle is exchanged among Electronic Control Units (ECUs) via in-vehicle network protocols such as the Controller Area Network (CAN) protocol.
	Since these protocols are designed for an isolated network, the protocols do not encrypt data nor authenticate messages.
	Intrusion Detection Systems (IDSs) are developed to secure the CAN protocol by detecting abnormal deviations in physical properties.
	For instance, a voltage-based IDS (VIDS) exploits voltage characteristics of each ECU to detect an intrusion.
	An ECU with VIDS must be connected to the CAN bus using extra wires to measure voltages of the CAN bus lines.
	These extra wires, however, may introduce new attack surfaces to the CAN bus if the ECU with VIDS is compromised.
	We investigate new vulnerabilities of VIDS and demonstrate that an adversary may damage an ECU with VIDS, block message transmission, and force an ECU to retransmit messages.
	In order to defend the CAN bus against these attacks, we propose two hardware-based Intrusion Response Systems (IRSs) that disconnect the compromised ECU from the CAN bus once these attacks are detected.
	We develop four voltage-based attacks by exploiting vulnerabilities of VIDS and evaluate the effectiveness of the proposed IRSs using a CAN bus testbed.
\end{abstract}


\IEEEpeerreviewmaketitle

\section{Introduction}
\label{sec:intro}

Electronic Control Units (ECUs) in a vehicle exchange data via in-vehicle network protocols such as Controller Area Network (CAN) \cite{ISO:2015}, Local Interconnect Network (LIN) \cite{LIN:2016}, and FlexRay \cite{FlexRay:2013}.
These in-vehicle network protocols do not have cryptographic primitives such as data encryption or message authentication because in-vehicle networks were designed to be isolated from external networks \cite{ISO:2015,LIN:2016,FlexRay:2013}.
Modern vehicles, however, are equipped with many ECUs that have outward-facing interfaces such as auxiliary port (AUX), Wi-Fi, cellular network, and Bluetooth.
These ECUs may introduce attack surfaces to the CAN bus as well as ECUs \cite{Miller:2015:remote,Miller:2013:adventure,miller2014survey,Koscher:2010:experimental,Checkoway:2011:comprehensive}.
It is difficult to upgrade the existing in-vehicle network protocols to encrypt data or authenticate messages due to the lack of backward compatibility with legacy systems and the resource constraints of ECUs such as memory size \cite{Murvay:VoltageIDS:2014}.
Hence, Intrusion Detection Systems (IDSs) have been developed to detect attacks on the CAN bus by tracking abnormal deviations in physical properties of the CAN bus or ECUs \cite{Shin:2016:finger,Shin:2016:error}.
Commonly exploited physical properties are message frequency \cite{Hoppe:2008:security}, clock skew of an ECU \cite{Shin:2016:finger,Sagong:2018:CCE:3207896.3207901}, entropy of the CAN bus \cite{Muter:2011:entropy}, and voltage levels of the CAN bus \cite{Cho:2017:VAI:3133956.3134001,Murvay:VoltageIDS:2014,Choi:VoltageIDS:2018}.

\begin{figure}[t!]
	\centering
	\includegraphics[width=0.45\textwidth]{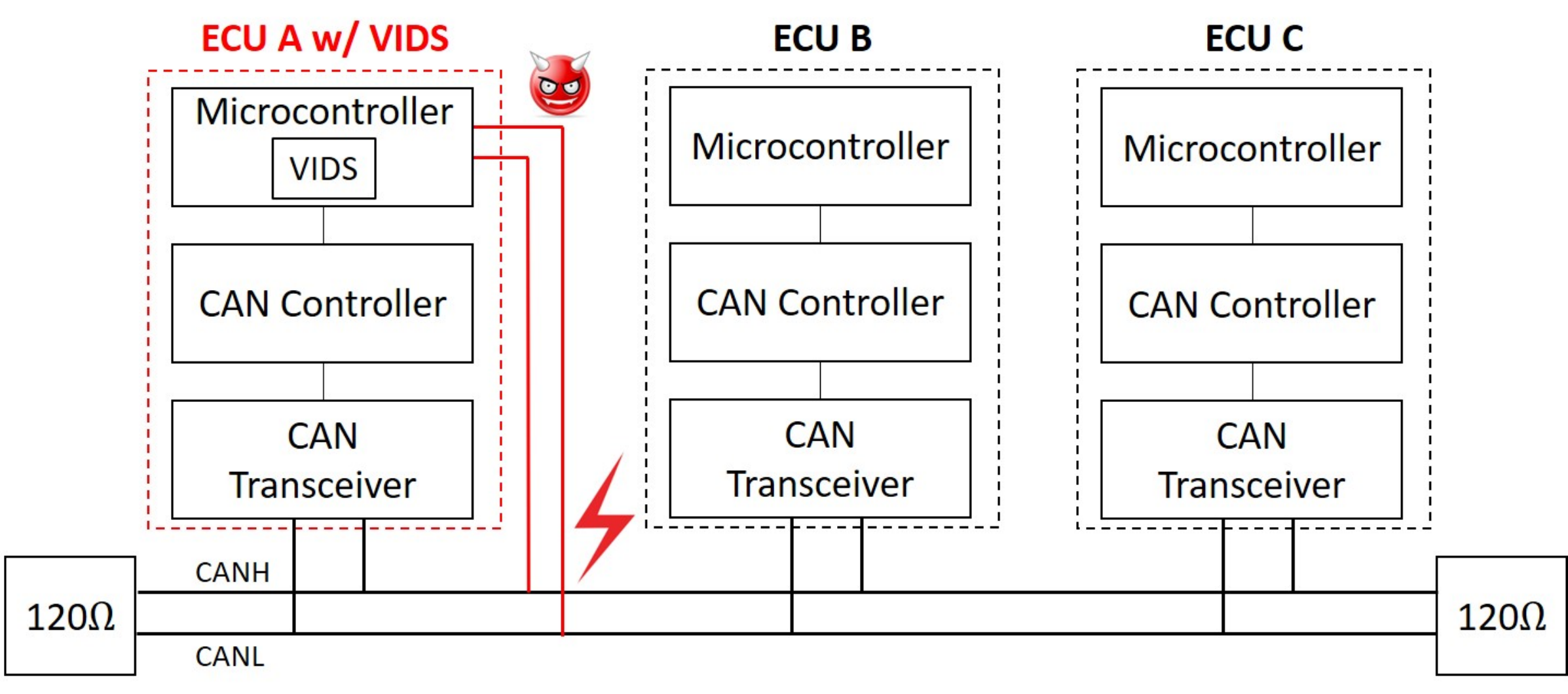}
	\caption{Architecture of VIDS.
	A VIDS is implemented on the microcontroller of ECU A.
	In order to measure voltages of the CAN bus lines, the microcontroller of ECU A is directly connected to the CAN bus lines via two extra wires (two red lines).
	If ECU A is compromised, the adversary may launch voltage-based attacks using the directly connected wires.}
	\label{fig:ECU_with_VIDS}
\end{figure}

It is difficult for an adversary to mimic voltage characteristics of an ECU because they depend on the CAN transceiver's hardware such as transistors and diodes.
As a result, using voltage characteristics as a fingerprint of an ECU is more effective and reliable than other physical properties, such as clock skew, in detecting attacks on the CAN bus \cite{Cho:2017:VAI:3133956.3134001,Murvay:VoltageIDS:2014,Choi:VoltageIDS:2018}.
A voltage-based IDS (VIDS) is implemented as software in the microcontroller of an ECU.
A CAN transceiver measures the voltage difference between the CAN bus lines (i.e., CANH and CANL), which is not sufficient to establish voltage characteristics.
Also, the microcontroller cannot access measured voltage values since a CAN controller only provides the decoded bit information to the microcontroller.
In order to measure voltages of the CAN bus lines, the microcontroller is directly connected to CANH and CANL using extra wires, respectively, as illustrated in Fig.~\ref{fig:ECU_with_VIDS}.
These two extra wires introduce new attack surfaces on the CAN bus and ECUs.
If an ECU installed with VIDS is compromised, the adversary may manipulate voltage levels of CANH and CANL using these wires in order to disable VIDS or impede CAN transceivers' operations.

In this paper, we propose four voltage-based attacks using the extra wires: 1) an overcurrent attack, 2) a denial-of-service (DoS) attack, 3) a forced retransmission attack, and 4) a pulse attack.
In the overcurrent attack, the adversary makes the current that flows into the microcontroller with VIDS exceed the hardware limit of the microcontroller (i.e., the current absolute maximum rating), which may damage the microcontroller. 
In the DoS attack, the adversary keeps the CAN bus in an idle state by holding the voltages of the CAN bus lines at one level, by which messages cannot be transmitted.
In the forced retransmission attack, the adversary violates the bit timing requirement of the CAN protocol to make an error, which makes a message be retransmitted.
In the pulse attack, the adversary arbitrarily changes the voltages of the CAN bus lines by applying a pulse signal, which blocks message transmission.

We make the following contributions in this paper:
\begin{itemize}
	\item We propose four voltage-based attacks that are an overcurrent attack, a DoS attack, a forced retransmission attack, and a pulse attack by manipulating the voltage levels of the CAN bus lines using the extra wires.
	
	\item We propose a fuse-based Intrusion Response System (IRS) and a heat-based IRS that detect and mitigate the voltage-based attacks by isolating the compromised ECU from the CAN bus.
		
	\item We demonstrate the proposed voltage-based attacks and evaluate the effectiveness of the proposed IRSs using a CAN bus testbed.
	Our evaluation shows that the voltage-based attacks can be implemented in practice.
	We also demonstrate that the proposed IRSs mitigate the voltage-based attacks.
\end{itemize}

The rest of the paper is organized as follows.
The related work on the IDSs for the CAN protocol is reviewed in Section \ref{sec:related}.
Section \ref{sec:preliminaries} summarizes a brief background on the CAN protocol and explains a CAN transceiver's operation.
Section \ref{sec:model} presents the adversary model, and Section \ref{sec:attack} describes the voltage-based attacks.
The hardware-based IRSs are proposed in Section \ref{sec:defense}.
The experimental results are presented in Section \ref{sec:eval}.
Section \ref{sec:conclusion} concludes the paper.

\section{Related Work}
\label{sec:related}

Many works have proposed various IDSs that detect cyber attacks using abnormal deviations in the traffic through the CAN bus.
Based on the fact that most of the messages in the CAN protocol are transmitted with a fixed length and frequency, an IDS that detects the existence of spoofed messages using a frequency of message occurrence is proposed \cite{Hoppe:2008:security}.
Also, the authors of \cite{Muter:2011:entropy} proposed the entropy-based IDS that exploits coincidence among a set of messages.
The entropy-based IDS, however, can be bypassed if an adversary replicates structure and pattern of the legitimate traffic \cite{Shin:2016:finger}.
Hence, IDSs that fingerprint each ECU by exploiting ECU's physical properties are developed \cite{Shin:2016:finger,Cho:2017:VAI:3133956.3134001,Choi:VoltageIDS:2018}.
For instance, the authors of \cite{Shin:2016:finger} proposed the clock-based IDS (CIDS) that uses the clock skew of ECUs to fingerprint each ECU.
It is demonstrated that the CIDS can be bypassed by the cloaking attack that matches the interarrival time of the spoofed messages to that of the legitimate messages \cite{Sagong:2018:CCE:3207896.3207901,cloaking_attack_tifs}.

The voltage characteristics are determined by the CAN transceiver's hardware and cannot be modified by the microcontroller's software, which makes mimicking the voltage characteristics through cyber attacks difficult for an adversary \cite{Cho:2017:VAI:3133956.3134001}.
Although connecting the microcontroller's analog pins to the CAN bus lines has a potential risk, many works have proposed VIDSs \cite{Cho:2017:VAI:3133956.3134001,Choi:VoltageIDS:2018,Murvay:VoltageIDS:2014}.
In \cite{Murvay:VoltageIDS:2014}, the proposed VIDS exploits the mean squared error between the measured voltage and each ECU's reference voltage that has been collected before an attack.
The VIDS proposed in \cite{Cho:2017:VAI:3133956.3134001} measures the voltages of each CAN bus line and extracts features from the distribution of the measured voltage samples.
The voltage difference between the CAN bus lines and transition time between bits 0 and 1 are exploited to detect abnormal deviations in the voltage characteristics by using machine learning algorithms in \cite{Choi:VoltageIDS:2018}.

VIDSs measure the voltages of CANH and CANL to extract the voltage characteristics of each ECU.
An oscilloscope is used to measure the voltage in \cite{Murvay:VoltageIDS:2014} and \cite{Choi:VoltageIDS:2018}, which may not be possible in a car due to limitations in power and space.
Alternatively, a VIDS may be implemented in a microcontroller, which requires 12V that can be supplied from a car battery \cite{Cho:2017:VAI:3133956.3134001}.
If the microcontroller or oscilloscope is compromised, the adversary may exploit the wires connected to the CAN bus lines to launch attacks.
The recent works on the VIDSs, however, did not discuss any potential cyber attacks in which these wires are maliciously used, nor did the works propose a protection mechanism to mitigate the attacks \cite{Cho:2017:VAI:3133956.3134001,Choi:VoltageIDS:2018,Murvay:VoltageIDS:2014}.
We consider such attacks in this paper.

Despite the effectiveness of an IDS in protecting the CAN bus, the IDS is limited to detecting cyber attacks and alerting an operator of the automobile.
The reaction to the detected attacks remains to the operator of the automobile \cite{carver:IRS:2000}.
An IRS, however, can detect an attack and promptly mitigate the attack \cite{carver:IRS:2000,Hoppe:2008:security}.

The contributions of this paper differ from our previous work \cite{Sagong:2018:escar} in that we investigate more attack surfaces of VIDS and improve a state-of-the-art IRS against the voltage-based attacks.
We develop a pulse attack that may block message transmission using pulse signals.
We also develop a heat-based IRS that can be reusable without replacing a fuse or resetting a circuit breaker after mitigating a voltage-based attack.

\section{Preliminaries}
\label{sec:preliminaries}

In this section, the CAN protocol is reviewed, and the operation of a CAN transceiver is discussed.
We explain analog pin modes of a microcontroller, which are used for measuring the voltage and generating electrical signals.

\subsection{CAN Protocol Background}

The CAN protocol is a multi-master broadcast bus network in which any ECU can transmit messages and receive all ongoing messages through the CAN bus.
An ECU that accesses the CAN bus first can transmit a message.
If two or more ECUs attempt to send messages simultaneously, the message with the smallest ID is first transmitted through a content-based collision detection process called \textit{arbitration}.
For example, consider two ECUs A and B that try to send their messages with IDs 0x010 and 0x001, respectively.
ECU A recognizes that a higher priority message is being transmitted and stops transmitting its message through an arbitration.
Bits 0 and 1 are also called dominant and recessive bits, respectively.

\begin{figure}[t!]
	\centering
	\includegraphics[width=0.45\textwidth]{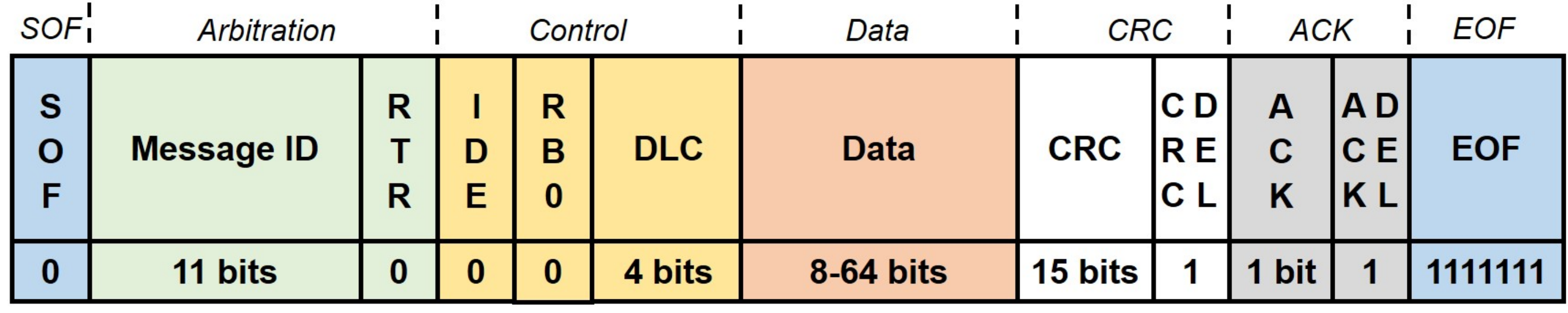}
	\caption{Structure of a data frame in the CAN protocol.}
	\label{fig:CAN_frame}
\end{figure}

A data frame in the CAN protocol consists of 7 fields as illustrated in Fig.~\ref{fig:CAN_frame}, and the length of the data field can be varied from 1 to 8 bytes.
As shown in the structure of the data frame, the data field is not encrypted, and there does not exist a field for message authentication.
If a message is not transmitted and received correctly because of external electromagnetic interference or malfunction of CAN transceivers, the ECU retransmits that message after an error frame is transmitted.

\subsection{Operation of CAN Transceivers}

\begin{figure}[t!]
	\centering
	\includegraphics[width=0.25\textwidth]{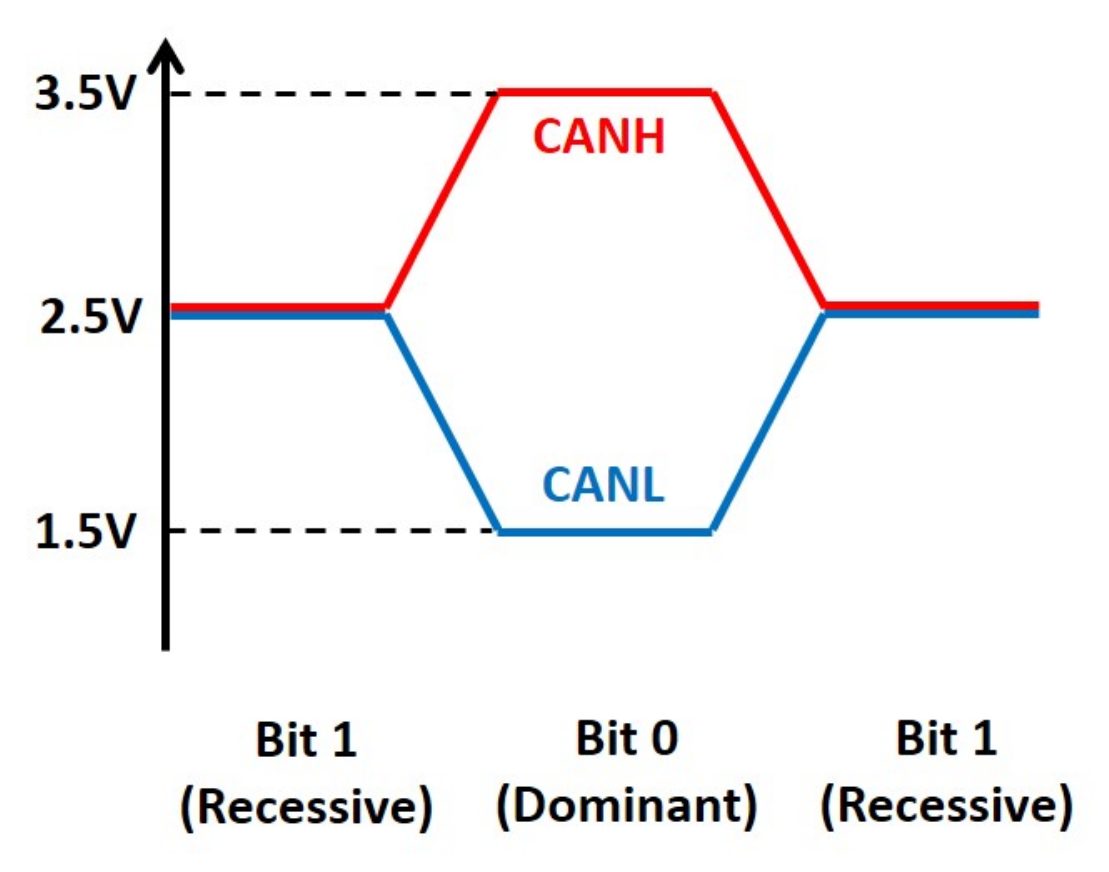}
	\caption{Voltage levels of CANH and CANL when dominant and recessive bits are transmitted by a CAN transceiver whose supply voltage is 5.0V.}
	\label{fig:voltage_level_5V}
\end{figure}

The CAN bus consists of two bus lines terminated by two 120$\Omega$ resistors.
In order to make the CAN bus more robust to the external electromagnetic interference, the CAN protocol uses the differential voltage between CANH and CANL to represent a bit.
When transmitting a recessive bit, both CANH and CANL are set to the same value, which makes the differential voltage be 0.0V.
When transmitting a dominant bit, CANH and CANL are set to different voltages to make the differential voltage larger than a predetermined threshold, typically 0.9V.
Fig.~\ref{fig:voltage_level_5V} illustrates the voltage levels of CANH and CANL when transmitting dominant and recessive bits using a CAN transceiver whose supply voltage is 5.0V.
For a recessive bit, both CANH and CANL are set to nominal 2.5V to make the differential voltage be 0.0V.
When transmitting a dominant bit, the differential voltage is 2.0V since CANH and CANL are set to nominal 3.5V and 1.5V, respectively.\footnote{For a recessive bit, the voltage levels of CANH and CANL are 2.3V when we use a CAN transceiver whose supply voltage is 3.3V, in which the differential voltage is still 0.0V. For a dominant bit, the CAN transceiver sets CANH and CANL to 3.0V and 1.0V, respectively. Hence, the differential voltage becomes 2.0V \cite{TexasInst:2013:CAN_voltage,TexasInst:2015:SN65HVD}. Although the supply voltage of the CAN transceivers might be different, the CAN transceivers exploit the differential voltage between CANH and CANL to represent a bit \cite{Microchip:MCP2551,NXP:TJA1043,TexasInst:2013:CAN_voltage,TexasInst:2015:SN65HVD,TexasInst:2016:CAN_tutorial,TexasInst:2008:CAN_Phy_Req}.}

\begin{figure}[t!]
	\centering
	\includegraphics[width=0.25\textwidth]{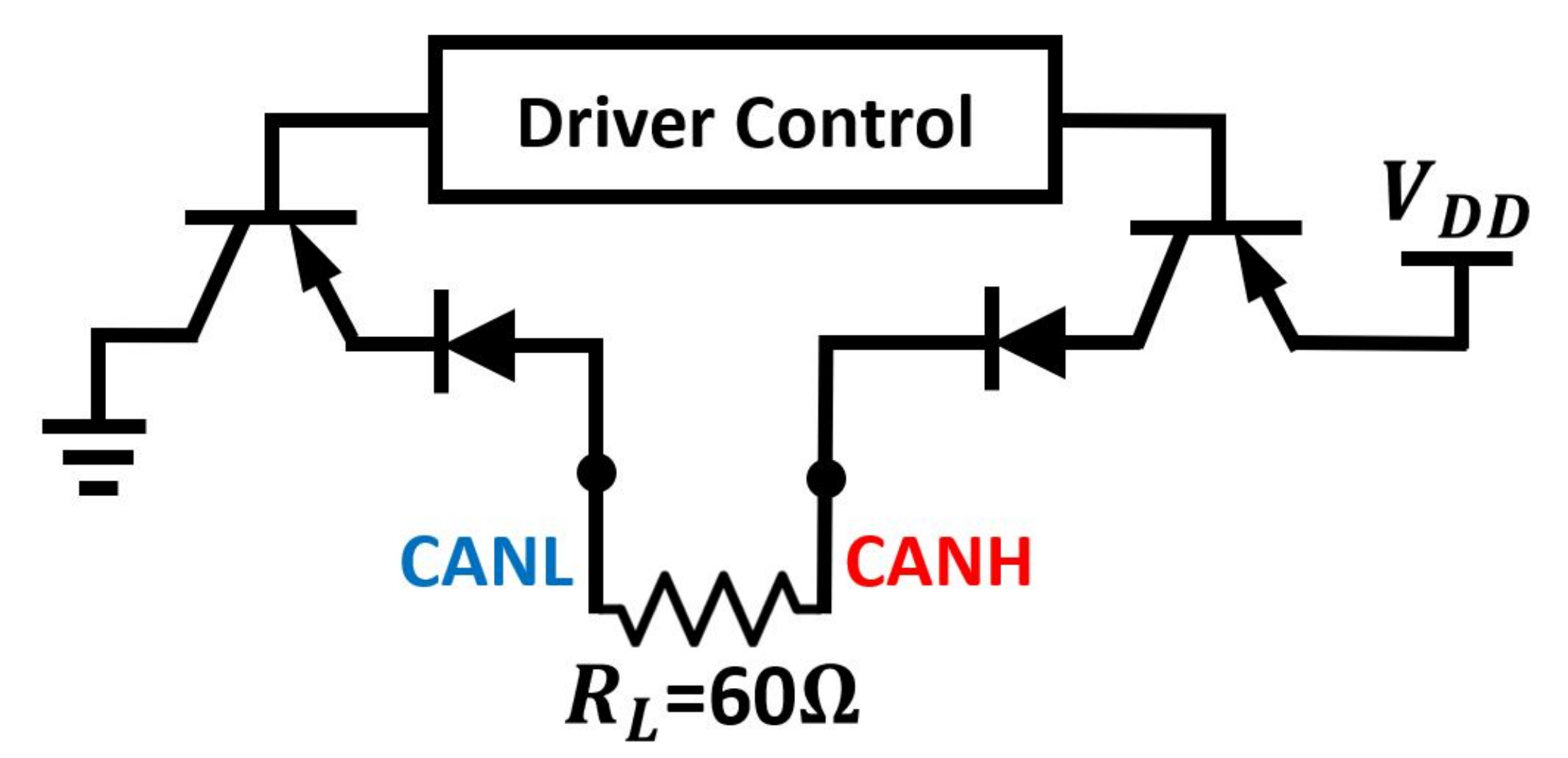}
	\caption{Schematic of Microchip MCP2551 CAN transceiver that is composed of two BJTs and two diodes. The driver control turns off the BJTs when transmitting a recessive bit and turns them on when transmitting a dominant bit. The impedance $R_L$ between CANH and CANL is 60$\Omega$ because two 120$\Omega$ termination resistors are connected in parallel.}
	\label{fig:transistor}
\end{figure}

The two transistors in a CAN transceiver control the voltage levels of CANH and CANL to transmit bits as illustrated in Fig.~\ref{fig:transistor}.
When transmitting a recessive bit, the transistors are in the off-state in which current cannot flow from the supply voltage $V_{DD}$ to the ground.
Due to the reference voltage (i.e., nominal 2.5V) applied to both CANH and CANL, the voltage levels of the CAN bus lines are set to the reference voltage.
As a result, the differential voltage becomes 0.0V that represents the recessive bit.

When transmitting a dominant bit, the driver control lets the transistors be in the on-state in which an electric path is created between the emitter and the collector in bipolar junction transistors (BJTs) or the source and the drain in field effect transistors (FETs).
Then, current flows from $V_{DD}$ to the ground.
Due to this current flowing through the termination resistors, the voltage is dropped from CANH to CANL, and the differential voltage becomes non-zero.
Since the operation of BJT and FET as a switch is identical, the analysis on the operation of a Microchip MCP2551 CAN transceiver that is composed of BJTs can be applied to any types of CAN transceivers that consist of FETs \cite{NXP:TJA1043}.
A Microchip MCP2551 CAN transceiver sets the voltages of CANH and CANL to 3.5V and 1.5V, respectively.
The voltage is dropped by 0.7V between the base and the emitter of the BJT when the BJT is in the on-state.
Since $V_{DD}$ is 5.0V, the voltage at the collector of the BJT connected to CANH becomes 4.3V.
The voltage is again dropped by 0.7V after the diode, which makes the voltage of CANH be 3.5-3.6V.
Similarly, the voltage of CANL becomes 1.4-1.5V because the voltage is increased by 0.7V at the BJT and the diode, respectively.

\subsection{Analog Pin Modes in Microcontroller}

A microprocessor measures voltage via its analog pin set to the input mode by using an Analog-to-Digital Converter (ADC) that is embedded in the microprocessor \cite{Arduino:UnoSpec,Renesas:V850,TexasInstruments:2016:arm_processor,Microchip:ATmega328PAuto,NXP:2005:MPC561}.
Also, a microprocessor may generate electric signals such as a Pulse Width Modulation (PWM) signal using an analog pin if that analog pin is set to the output mode \cite{Arduino:UnoSpec,Renesas:V850,TexasInstruments:2016:arm_processor}.
For example, Microchip ATmega328P used in an Arduino UNO Rev3 board, NXP MPC563 used in an ECU of Hyundai Sonata, and Renesas V850 used in an Engine Control Module of Toyota Camry provide analog pins that can be used to measure voltage or generate PWM signals \cite{Microchip:ATmega328PAuto,NXP:2005:MPC561,Renesas:V850}.

\section{Adversary Model}
\label{sec:model}

In this section, we describe an adversary model for voltage-based attacks on the CAN bus and VIDS.
Consider first how an ECU with VIDS can be compromised.
If an adversary accesses the CAN bus physically via an On-Board Diagnostics (OBD)-II port that is mandated for all cars in the US\cite{US:OBD} and EU\cite{EU:OBD}, the adversary may upload its malicious code to the ECU with VIDS using a pass-thru device such as Hyundai Global Diagnostic System \cite{Hyundai:GDS}, Ford Vehicle Communication Module \cite{Ford:VCM}, Volkswagen VAG-COM Diagnostic System \cite{Volkswagen:VCDS}, or Toyota Technical Information System \cite{Toyota:TIS}.
The adversary may also remotely compromise an ECU without physical access to the CAN bus if that ECU is equipped with a telematics unit \cite{Miller:2015:remote,Checkoway:2011:comprehensive,Rieke:Behavior:2017}.
Then, any ECU on the CAN bus including the ECU with VIDS can be compromised using the compromised ECU \cite{Checkoway:2011:comprehensive}.

Although an adversary may compromise an ECU with VIDS, the firmware of the CAN controller cannot be modified without a special equipment \cite{Atmel:2016:programmer}.
Hence, the adversary cannot manipulate the output voltage of the CAN transceiver in order to violate the CAN protocol.
The adversary, however, can manipulate the analog pin modes of the microcontroller once the ECU with VIDS is compromised.
As a result, the adversary may apply electric signals to the CAN bus by controlling the analog pin modes as explained in Section~\ref{sec:preliminaries}.

\section{Voltage-based Attacks}
\label{sec:attack}

\begin{table}[t!]
	\footnotesize
	\centering
	\caption{Combinations of analog pin modes of $P_H$ and $P_L$ for measuring the voltage and launching attacks.}
	\begin{tabular}{| c | c| c |}
		\hline
		$P_H$ mode
		& $P_L$ mode
		& Type of attack
		\\
		\hline
		Input & Input & Not an attack, setting for measuring voltage \\
		\hline
		Input & High & DoS attack \\
		\hline
		Input & Low & Passive overcurrent attack \\
		\hline
		High & Input & Forced retransmission attack \\
		\hline
		High & Low & Active overcurrent attack \\
		\hline
		Low & Input & DoS attack or passive overcurrent attack \\
		\hline
		Low & High & DoS attack or active overcurrent attack \\
		\hline
		Low & Low & DoS attack or passive overcurrent attack\\
		\hline
		Pulse & Input & Pulse attack\\
		\hline
		Input & Pulse & Pulse attack\\
		\hline
	\end{tabular}
	\label{table:attack_scenario}
	\normalsize
\end{table}

In this section, we introduce four proposed voltage-based attacks that are an overcurrent attack, a DoS attack, a forced retransmission attack, and a pulse attack.
Let us denote two analog pins of the microcontroller with VIDS that are connected to CANH and CANL as $P_H$ and $P_L$, respectively.
We discuss combinations of $P_H$ and $P_L$ for the normal operation of a VIDS and voltage-based attacks.
An analog pin can be in either 1) input for measuring the voltage, 2) high voltage output, 3) low voltage output, or 4) pulse output.
We consider two levels of the voltage output, and a PWM signal can be generated by switching low and high voltage output modes in the pulse output mode.\footnote{Depending on microcontrollers, an analog pin may generate multiple levels of the output voltage. In this paper, we consider the binary levels of the output voltage \cite{Arduino:UnoSpec}.}

In the normal operation of a VIDS, both $P_H$ and $P_L$ are in the input mode to measure the voltage levels of CANH and CANL, respectively.
If the adversary compromises an ECU with VIDS, the adversary may change the pin modes of $P_H$ and $P_L$ by uploading its malicious code.
Some combinations of the analog pin modes may damage the microcontroller with VIDS, block message transmission, or cause message retransmission as listed in Table~\ref{table:attack_scenario}.

\subsection{Overcurrent Attack}

A microcontroller may absorb current below a threshold through an analog pin due to its hardware limitation, which is called the current absolute maximum rating, $I_{max}$.
If current larger than this limit flows into the analog pin, the microcontroller can be damaged by an electric shock.
The idea of the overcurrent attack is that the adversary makes the current larger than $I_{max}$ flow into $P_L$.
For instance, values of $I_{max}$ are 40mA for Microchip ATmega328P \cite{Arduino:UnoSpec} and Renesas V850 \cite{Renesas:V850:Datasheet} and 20mA for NXP MPC563 \cite{NXP:2005:MPC561}.\footnote{The datasheet of Renesas V850 does not provide the current absolute maximum rating. Since the voltage absolute maximum rating of Renesas V850 and operating condition such as temperature are similar to that of Microchip ATmega328P, it is reasonable to assume that the current absolute maximum rating of Renesas V850 is similar to Microchip ATmega328P, which is 40mA.}

A small current flows through the analog pin of the microcontroller when measuring voltage.
Due to the high impedance at the microcontroller, current drawn by the microcontroller is negligible, which is in the order of nA.
The current flowing through an analog pin during the voltage measurement is much smaller than $I_{max}$ and does not damage the microcontroller.

The adversary, however, may make the microcontroller absorb current through $P_L$ by manipulating the analog pin mode.
Let $V_{H,b}$ and $V_{L,b}$ be the voltages of CANH and CANL when bit $b$ is transmitted, respectively, where $b$ is either 0 or 1.
For instance, the voltage of CANL when a recessive bit is transmitted is denoted as $V_{L,1}$.
In order to let the current larger than $I_{max}$ flow into the microcontroller, $V_{H,0}-V_{L,0}$ must meet the following condition.
\begin{equation}
\frac{(V_{H,0}-V_{L,0})}{R_L} > I_{max},
\label{eq:sink_current}
\end{equation}
where $R_L$ is 60$\Omega$ and $I_{max}$ depends on a hardware limit of the microcontroller, typically 40mA.

\begin{figure}[t!]
	\centering
	\subfloat[Passive overcurrent attack\label{fig:concept_passive_overcurrent_circuit}] {\includegraphics[width=0.22\textwidth]{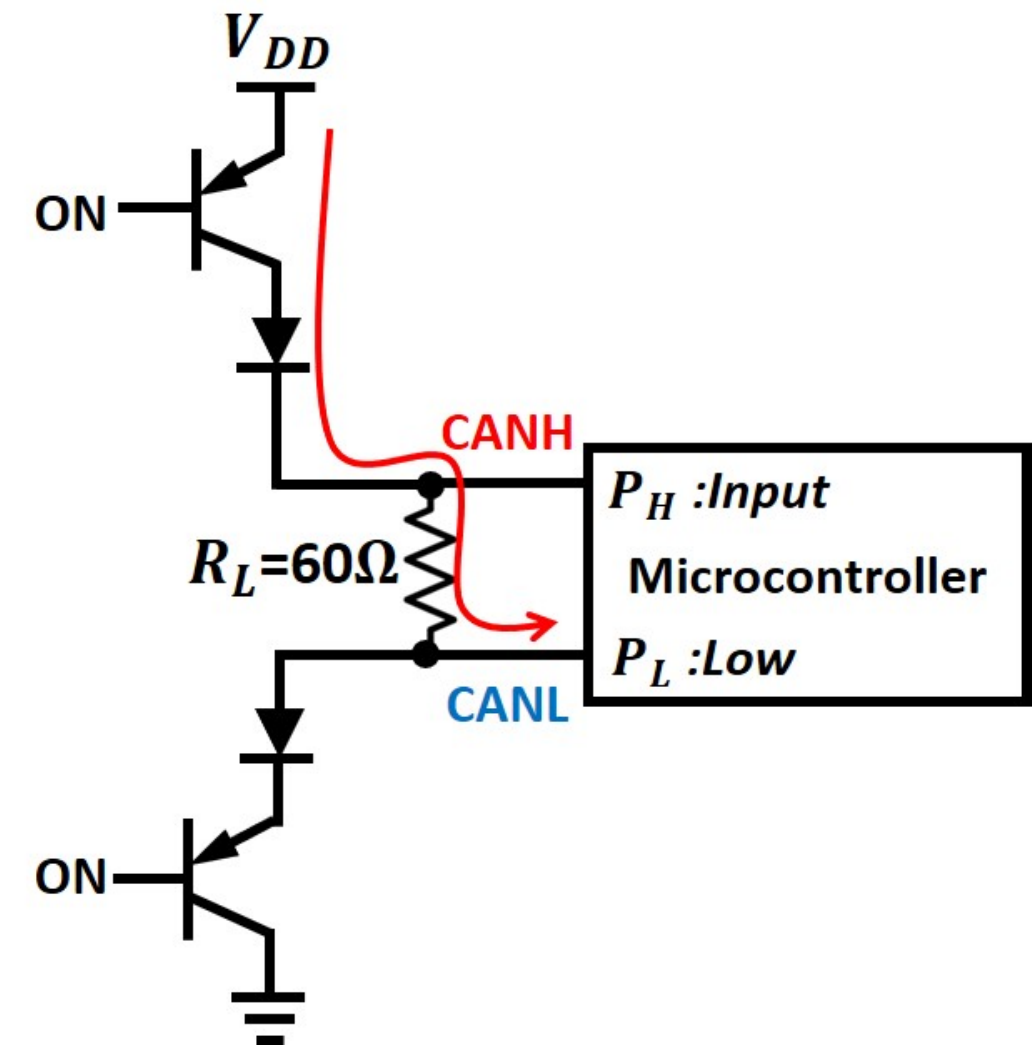}}
	\subfloat[Active overcurrent attack\label{fig:concept_active_overcurrent_circuit}]{\includegraphics[width=0.22\textwidth]{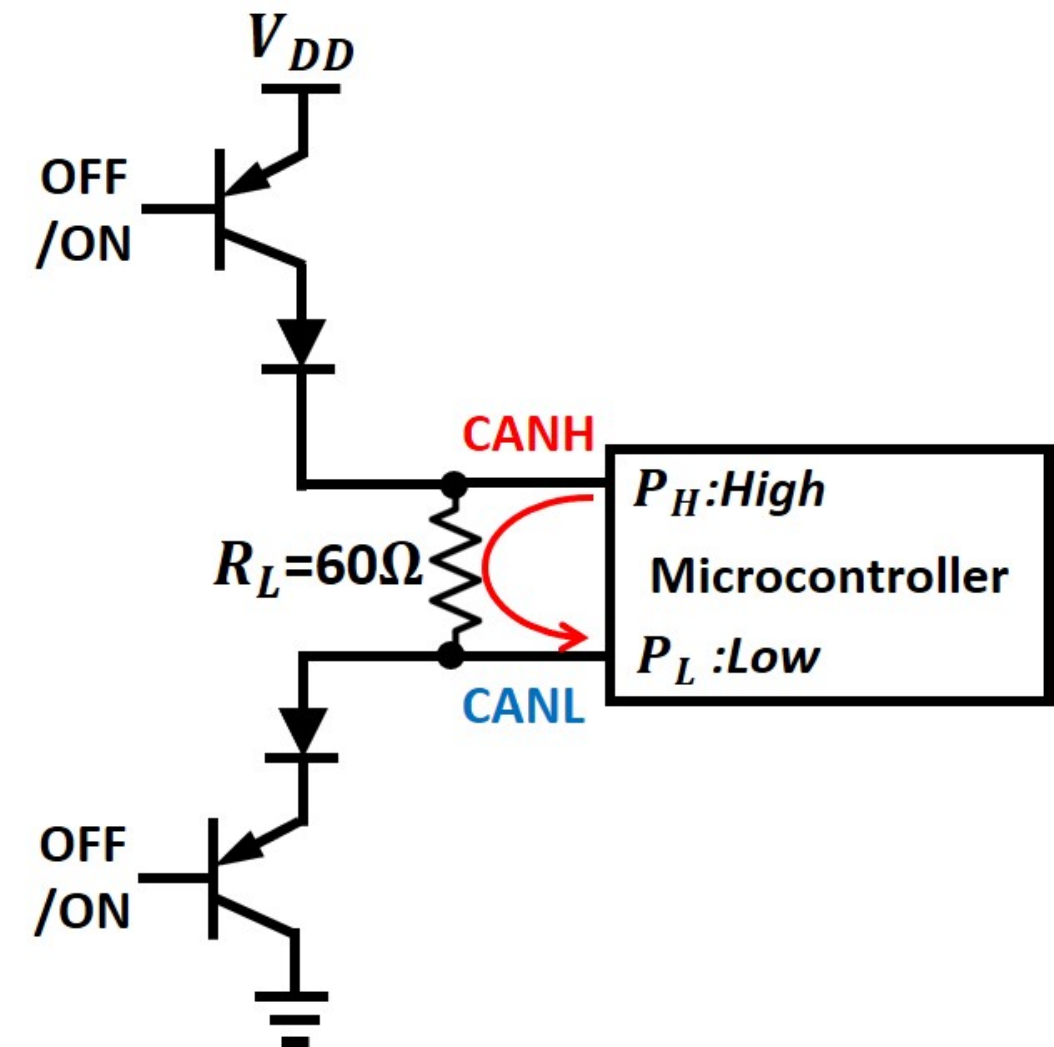}}
	\caption{Circuit diagrams under the overcurrent attacks. The red curve indicates the flow of the current in each overcurrent attack. (a) In the passive overcurrent attack, the current flows from $V_{DD}$ to $P_L$ when a dominant bit is transmitted. (b) In the active overcurrent attack, the current flows from $P_H$ to $P_L$ regardless of the bits.}
	\label{fig:concept_overcurrent}
\end{figure}

\begin{figure}[t!]
	\centering
	\subfloat[Passive overcurrent attack\label{fig:concept_passive_overcurrent_voltage}] {\includegraphics[width=0.23\textwidth]{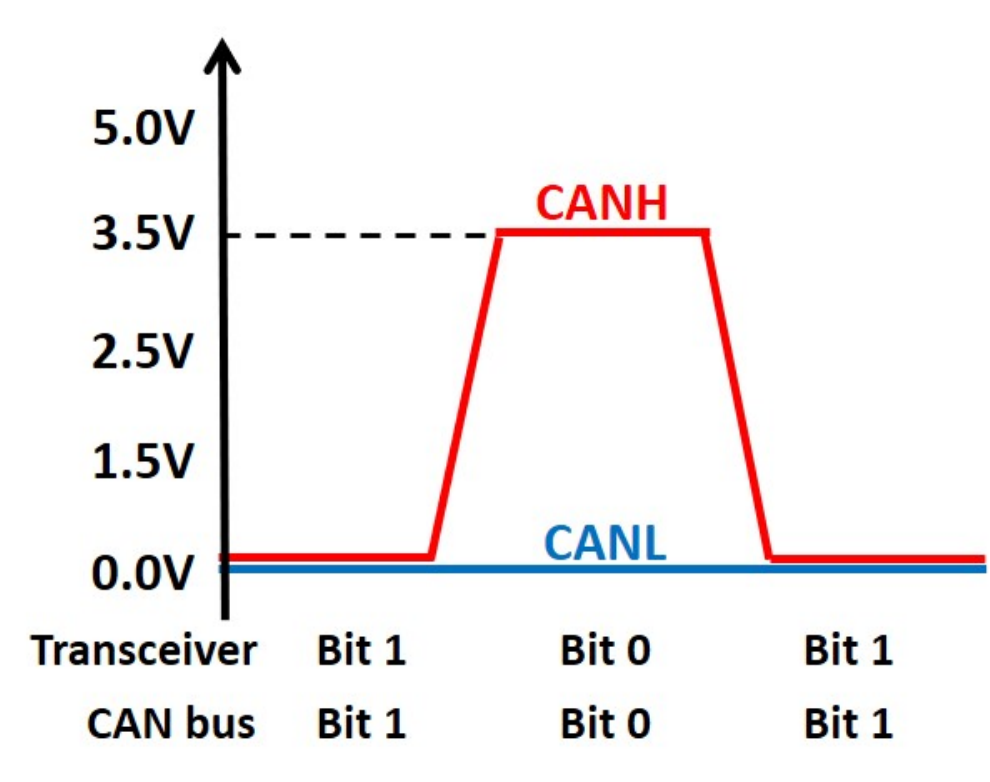}}
	\subfloat[Active overcurrent attack\label{fig:concept_active_overcurrent_voltage}]{\includegraphics[width=0.23\textwidth]{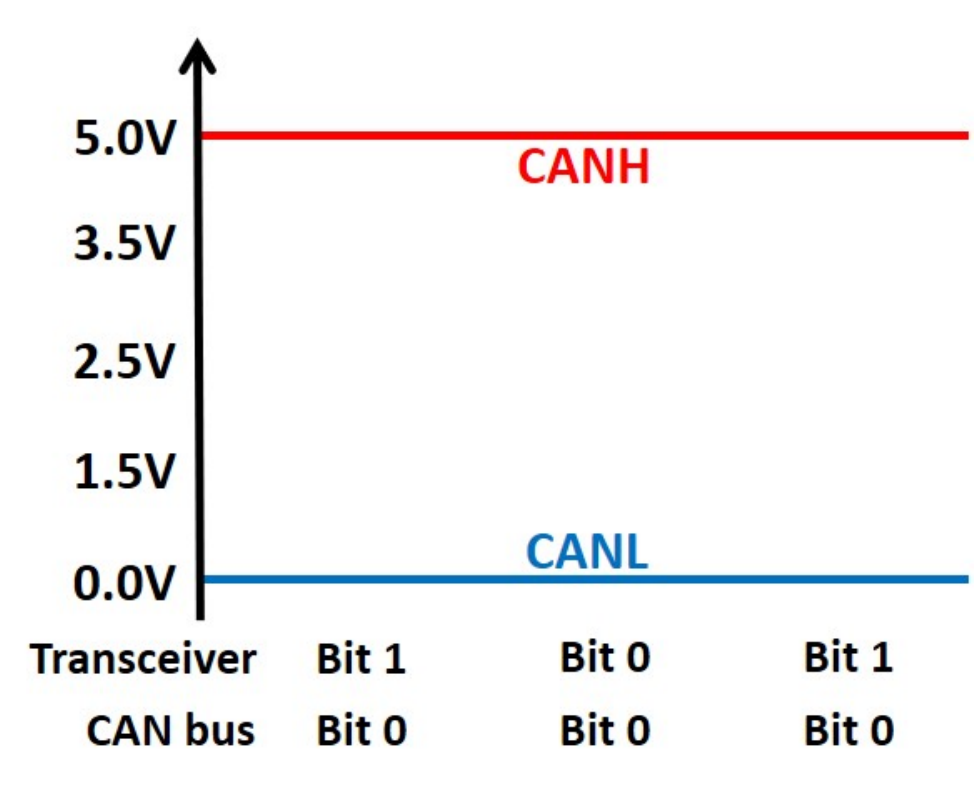}}
	\caption{Voltages of CANH and CANL under the overcurrent attacks. (a) In the passive overcurrent attack, the maximum voltage difference between CANH and CANL is 3.5V when a dominant bit is transmitted. (b) In the active overcurrent attack, the voltage difference is always 5.0V.}
	\label{fig:concept_overcurrent_voltage}
\end{figure}

We propose two types of the overcurrent attacks, namely passive and active overcurrent attacks, by changing the pin modes of $P_L$ and $P_H$ as illustrated in Fig.~\ref{fig:concept_overcurrent}.
The adversary only changes $P_L$ from the input mode to the low voltage output mode to launch the passive overcurrent attack.
Since $V_{H,0}$ is 3.5V as shown in Fig.~\ref{fig:concept_passive_overcurrent_voltage}, the current absorbed through $P_L$ can be computed as $\frac{3.5V-0.0V}{60\Omega}$=58.3mA, which satisfies Eq.~(\ref{eq:sink_current}).
In the passive overcurrent attack, the current is supplied from a car battery that can provide the current in the order of Ampere as illustrated in Fig.~\ref{fig:concept_passive_overcurrent_circuit}.
In order to launch the active overcurrent attack, the adversary changes $P_H$ and $P_L$ to the high voltage output mode and the low voltage output mode, respectively.
$V_{H,0}$ and $V_{L,0}$ become 5.0V and 0.0V, respectively, as shown in Fig.~\ref{fig:concept_active_overcurrent_voltage}.
Then, the current flowing through $P_L$ is computed as $\frac{5.0V-0.0V}{60\Omega}$=83.3mA that is also larger than $I_{max}$.
In the active overcurrent attack, the current is supplied from the microcontroller as illustrated in Fig.~\ref{fig:concept_active_overcurrent_circuit}.

\subsection{Denial-of-Service (DoS) Attack}

In order to transmit a bit on the CAN bus, the differential voltage between CANH and CANL, $V_{Diff}$, must satisfy the CAN protocol.
The idea of the DoS attack is that the adversary increases the voltage level of CANL such that $V_{Diff}$ is always smaller than the decision threshold for determining a dominant bit.
As a result, a recessive bit can only be transmitted, which means that the CAN bus is in the idle state.
Also, messages cannot be transmitted.
The adversary sets $P_L$ from the input mode to the high voltage output mode.
Let $V_{attack,L}$ denote the voltage applied to CANL by the adversary using $P_L$.
Then, $V_{Diff}$ can be computed as follows
\begin{equation}
V_{Diff} = V_{H,b} - V_{attack,L}.
\label{eq:diff_voltage}
\end{equation}

\begin{figure}[t!]
	\centering
	\includegraphics[width=0.23\textwidth]{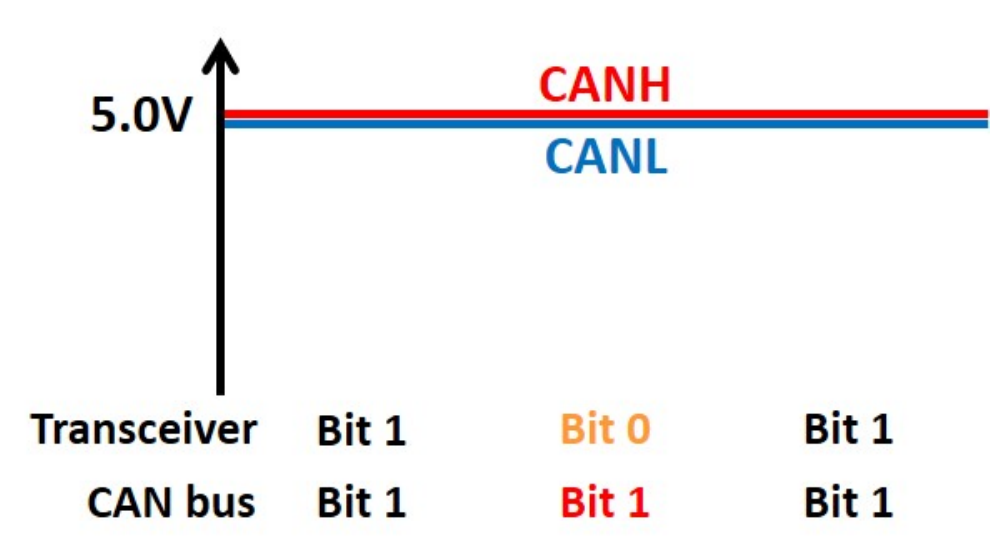}
	\caption{Voltages of the CAN bus under the DoS attack when $V_{attack,L}$ is 5.0V. Both CANH and CANL become 5.0V, which represents a recessive bit.}
	\label{fig:concept_voltage_DoS_attack}
\end{figure}

\begin{figure}[t!]
	\centering
	\includegraphics[width=0.38\textwidth]{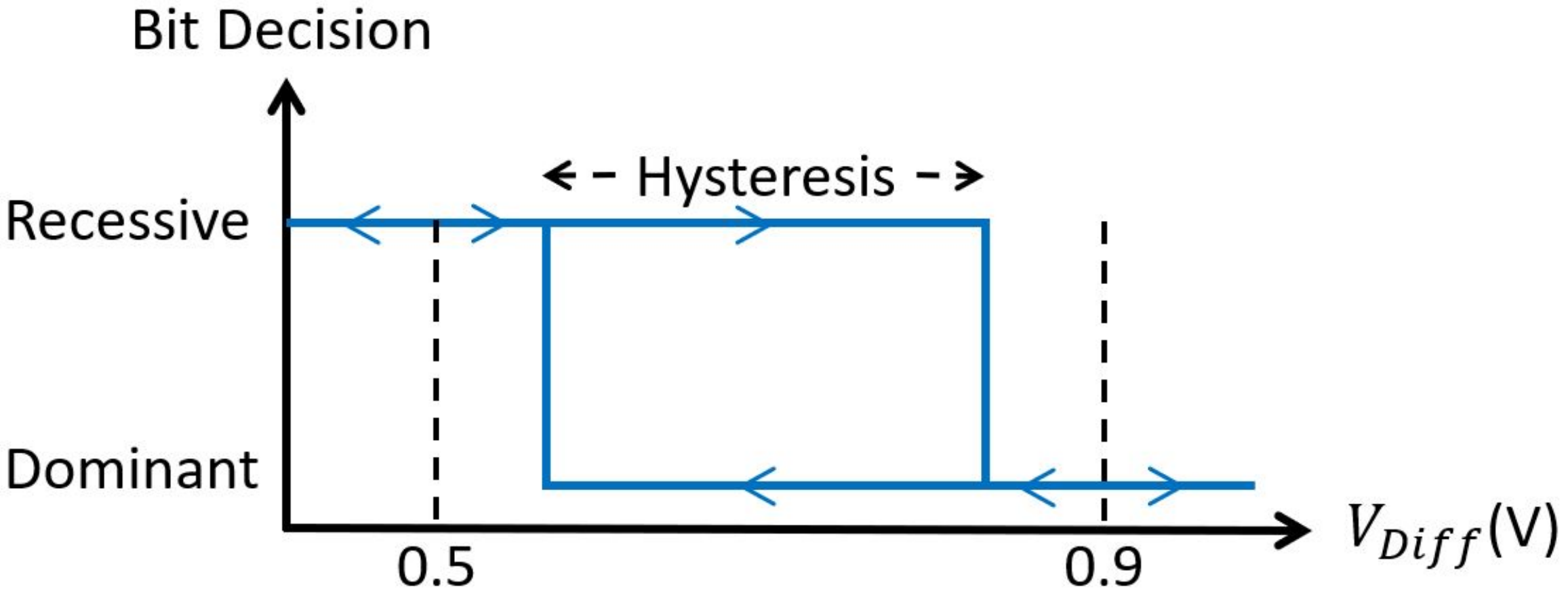}
	\caption{Bit decision criteria of Microchip MCP2551 CAN transceiver. A CAN transceiver can determine a dominant bit if $V_{Diff}$ is larger than 0.9V and a recessive bit if $V_{Diff}$ is less than 0.5V.}
	\label{fig:bit_decision}
\end{figure}

When a recessive bit is transmitted, both CANH and CANL are set to $V_{attack,L}$ under the DoS attack because current does not flow through the termination resistors.
Using Eq.~(\ref{eq:diff_voltage}), $V_{Diff}$ becomes 0.0V, which represents a recessive bit.
Although the voltages of CANH and CANL are manipulated as illustrated in Fig.~\ref{fig:concept_voltage_DoS_attack}, the recessive bit can be transmitted under the attack.
A CAN transceiver determines a dominant bit if $0.9$V$<$$V_{Diff}$$<$$5.0$V with the differential input hysteresis between 100-200mV as shown in Fig.~\ref{fig:bit_decision}.
When transmitting a dominant bit, the adversary may thwart the transmission of the dominant bit by setting $V_{attack,L}$ to make $V_{Diff}$ smaller than 0.9V.
If $V_{attack,L}$ is larger than 3.5V, $V_{H,0}$ also becomes $V_{attack,L}$ because current cannot flow through the termination resistors due to the diode in the CAN transceivers.
Hence, $V_{Diff}$ again becomes 0.0V, which indicates that the DoS attack is successfully launched.
Also, the adversary may launch the DoS attack by setting $P_H$ in the low voltage output mode since $V_{Diff}$ is always equal to or less than 0.0V.

\subsection{Forced Retransmission Attack}

The duration of one bit must satisfy the timing requirement of the CAN protocol according to the CAN bus speed.
The idea of the forced retransmission attack is that the adversary makes the transition time from a dominant bit to a recessive bit longer than the nominal value, typically 70-130ns \cite{Microchip:MCP2551}.
As a result, an error occurs while receiving a message, especially at the ACK delimiter position of a data frame as shown in Fig~\ref{fig:CAN_frame}.
The adversary increases the transition time, thus making a dominant bit at the ACK delimiter position that has to be a recessive bit.
The adversary changes $P_H$ from the input mode to the high voltage output mode to launch the forced retransmission attack.

\begin{figure}[t!]
	\centering
	\includegraphics[width=0.45\textwidth]{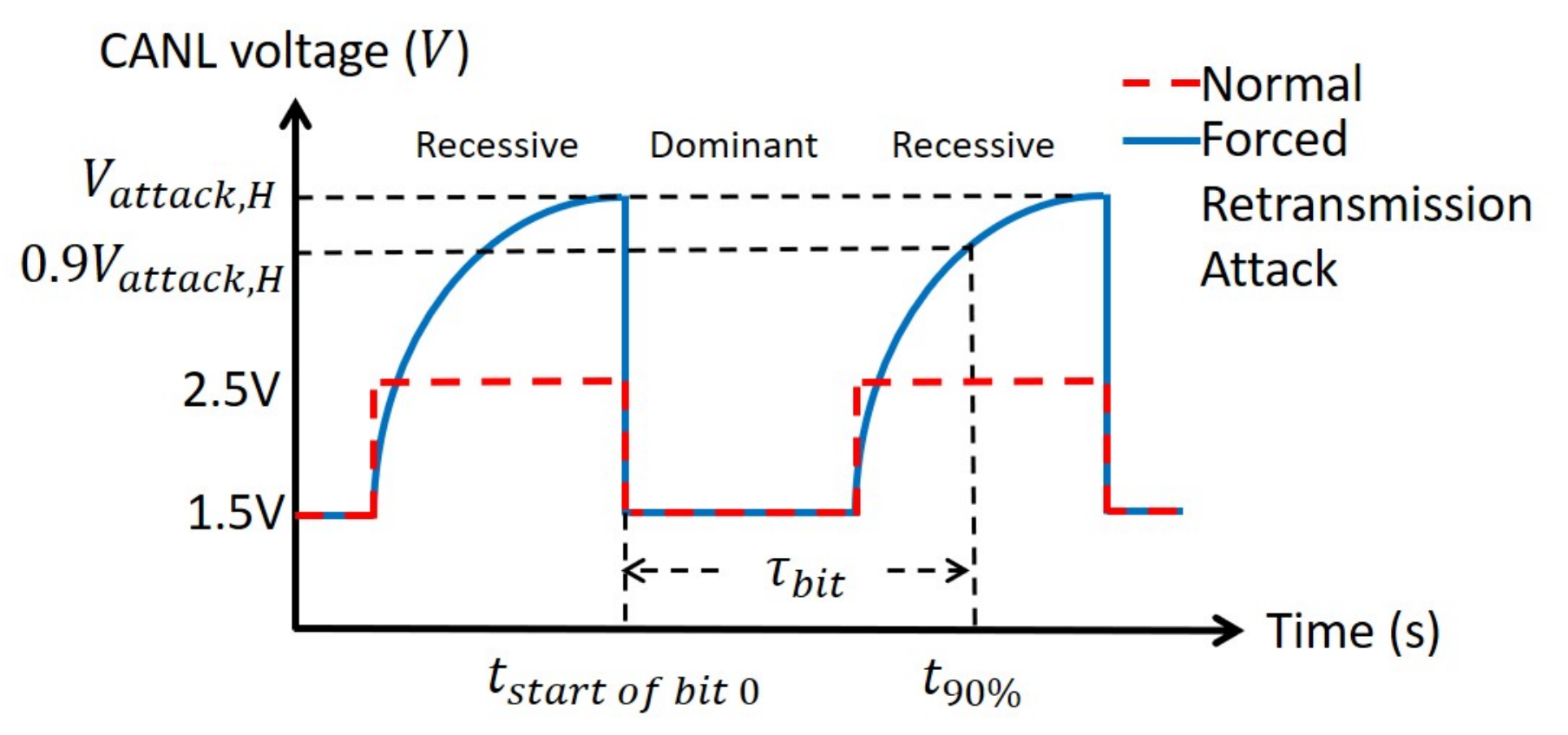}
	\caption{Voltage of CANL under the forced retransmission attack. Compared with the voltage level in the normal operation, the voltage of CANL is pulled up to $V_{attack,H}$, which increases the transition time. }
	\label{fig:bit_length_time}
\end{figure}

In order to analyze the transition time from a dominant bit to a recessive bit quantitatively, we define the \emph{bit length time} $\tau_{bit}$ as follows
\begin{equation}
\tau_{bit} \triangleq t_{90\%}-t_{start\,\,of\,\,bit\,\,0},
\label{eq:transition_time}
\end{equation}
where $t_{start\,\,of\,\,bit\,\,0}$ and $t_{90\%}$ denote the time at which a dominant bit starts and the time at which the voltage of CANL reaches 90\% of $V_{attack,H}$, respectively, as illustrated in Fig. \ref{fig:bit_length_time}.
$\tau_{bit}$ also indicates the sum of the duration of the dominant bit and the transition time.
For instance, if the CAN bus speed is set to 500kbps, $\tau_{bit}$ without the attack is nominal 2$\mu$s.

\begin{figure}[t!]
	\centering
	\includegraphics[width=0.23\textwidth]{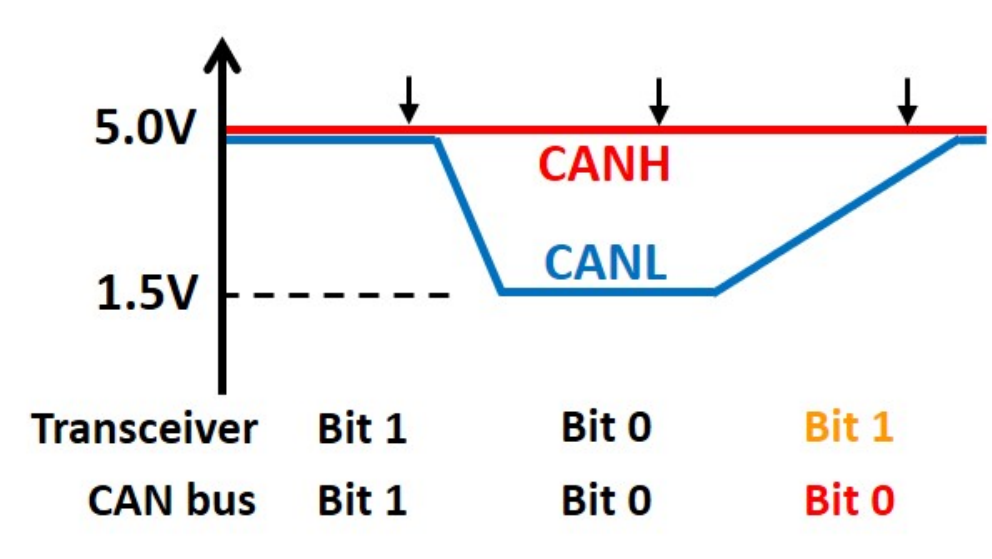}
	\caption{Voltages of the CAN bus under the forced retransmission attack when $V_{attack,H}$ is 5.0V. When a recessive bit is transmitted right after a dominant bit, the transition time of CANL's voltage level increases. Hence, an ECU may receive a dominant bit because $V_{Diff}$ is greater than 0.9V when sampling the voltages of the CAN bus lines at the instances indicated with the black arrows facing downward.}
	\label{fig:concept_voltage_FRA}
\end{figure}

Let $V_{attack,H}$ denote the voltage applied to CANH by the adversary via $P_H$, and we consider $V_{attack,H}$$>$2.5V.
Fig.~\ref{fig:concept_voltage_FRA} illustrates the voltages of CANH and CANL under the forced retransmission attack when $V_{attack,H}$ is 5.0V.
When transmitting a recessive bit, both CANH and CANL become $V_{attack,H}$ since current does not flow through the termination resistors.
A recessive bit can be transmitted under the attack.
A dominant bit can also be transmitted because $V_{Diff}$ is greater than 0.9V where $V_{Diff}$ increases as increasing $V_{attack,H}$.
If $V_{attack,H}$ becomes greater than 3.5V, the adversary forcefully sets CANL higher than the normal operating range of the CAN transceiver.
As a result, the voltage of CANL switches from 1.5V to $V_{attack,H}$ when a recessive bit is transmitted right after a dominant bit, which increases the transition time as illustrated in Fig.~\ref{fig:bit_length_time}.
Using its local clock, an ECU decodes a bit by sampling the voltage levels of CANH and CANL with the fixed period that is determined by the CAN bus speed.
If the duration of a bit is longer than the nominal value, the ECU cannot decode the bit correctly.
Hence, an error occurs while receiving a message, and that message is retransmitted.

\subsection{Pulse Attack}

The idea of the pulse attack is blocking message transmission using a PWM signal.
Consider that the PWM signal is applied to CANL using $P_L$.
Then, there are four possible cases for the voltage behavior of the CAN bus because the PWM signal could be in either the high voltage output mode or the low voltage output mode when a dominant or recessive bit is transmitted, respectively.
If the PWM signal is in the high voltage output mode while a dominant bit is transmitted, both CANH and CANL become the same voltage, which represents a recessive bit.
Hence, the dominant bit cannot be transmitted through the CAN bus, and message transmission is blocked.
A bit can be transmitted correctly in the other three cases because $V_{Diff}$ satisfies the CAN protocol.

A PWM signal and the start of the message are not perfectly synchronized in most of the time because the PWM signal is applied in arbitrary time without considering message transmission.
As a result, the case in which a dominant bit is tried to be transmitted when the PWM signal is in the high voltage output always occurs.
If the PWM signal is applied to CANH, a dominant bit cannot be transmitted if the PWM signal is in the low voltage output.
The pulse attack occurs when a PWM signal is applied to either $P_H$ or $P_L$ where the other analog pin is in the input mode.

A CAN controller determines a received bit using the voltage signal from the CAN transceiver.
In order to determine the bit, this voltage signal has to stay constant longer than a predetermined threshold, typically 300-350ns \cite{Microchip:MCP2515}.
Otherwise, the CAN controller cannot decode the received bit.
The impact of the PWM signal during message transmission can be negligible if the duration of blocking the transmission of a dominant bit is shorter than this predetermined threshold.
For example, when the duty cycle is fixed at 50\%, the period of the PWM signal has to be longer than 700ns to thwart the transmission of a dominant bit, which launches the pulse attack successfully.
If the period of the PWM signal is twice longer than the message transmission time for the fixed duty cycle 50\%, the PWM signal changes the voltage levels of the CAN bus lines for the entire message transmission time, which is identical to the DoS attack and forced retransmission attack.
Hence, we only consider the period less than the message transmission time in this paper.

\section{Hardware-based Intrusion Response Systems}
\label{sec:defense}

\begin{figure}[t!]
	\centering
	\subfloat[Fuse-based IRS\label{fig:structure_IRS}]{\includegraphics[width=0.25\textwidth]{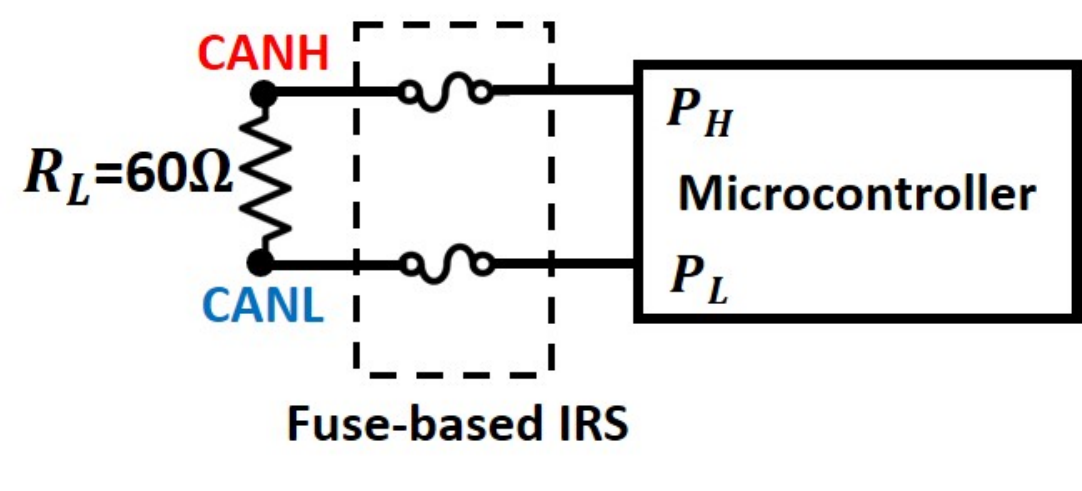}}
	\hfill
	\\
	\subfloat[Heat-based IRS\label{fig:structure_heat_IRS}]{\includegraphics[width=0.30\textwidth]{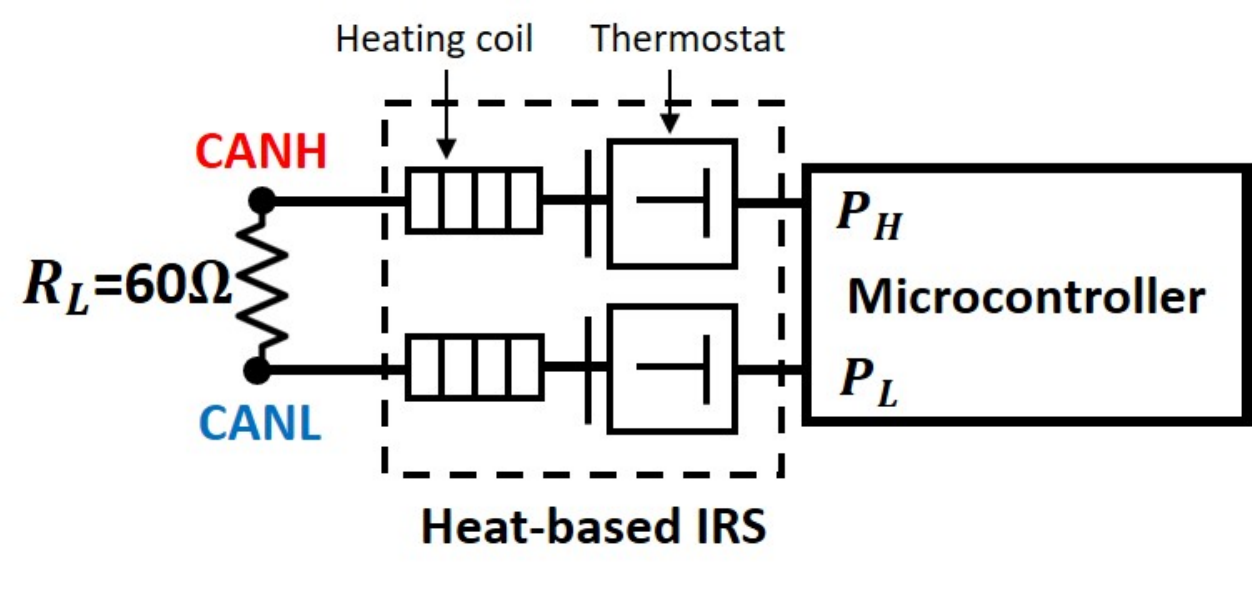}}
	\caption{Structures of the proposed hardware-based IRSs.}
	\label{fig:hardware_IRS}
\end{figure}

In this section, we propose two hardware-based IRSs, namely fuse-based IRS and heat-based IRS, and explain how these IRSs may mitigate the voltage-based attacks.
As shown in Fig.~\ref{fig:structure_IRS}, the fuse-based IRS consists of fuses or circuit breakers that are attached between the microcontroller's analog pins and the CAN bus lines, while the heat-based IRS consists of heating coil and thermostats as illustrated in Fig.~\ref{fig:structure_heat_IRS}.
The current flowing through the analog pins can be used as an indicator of the voltage-based attacks because the current flows through these analog pins under the voltage-based attacks. 
The hardware-based IRSs physically isolate a VIDS from the CAN bus as soon as any one of the voltage-based attacks is detected.
The hardware components such as fuses, circuit breakers, heating coil, or thermostats operate independently of the microcontroller.
Hence, the cyber attack on an automobile cannot disable the proposed IRSs.
In the rest of this section, we explain how each proposed IRS can mitigate the voltage-based attacks.

\subsection{IRSs for Overcurrent Attack}

The overcurrent attack can be mitigated by limiting the current that flows into the microcontroller to be less than the current absolute maximum rating, $I_{max}$.
A fuse may limit the current since it disconnects two parts of a circuit when current larger than its current rating flows longer than its opening time.
The typical opening time of very fast-acting fuses is in the order of $\mu$s to ms \cite{LittelFuse:272,LittelFuse:guide}.
Since it takes a longer time to damage a microcontroller by the overcurrent than to open a fuse \cite{Arduino:UnoSpec}, the fuse disconnects the wire before the microcontroller is burned.
The microcontrollers may be damaged by the overcurrent within a short amount of time in the order of ns and $\mu$s.
Micro-electro-mechanical system (MEMS)-based fuses or thin film-based fuses can be used for the fuse-based IRS because their opening times are in the order of $\mu$s for the MEMS-based fuse \cite{Chiu_fuse2005} and in the order of ns for the thin-film based fuse \cite{Ding_fuse2016}.
A fuse does not thwart the voltage measurement at the microcontroller when the fuse is attached to an analog pin as illustrated in Fig.~\ref{fig:structure_IRS} since the fuse is a wire that does not induce any voltage drop ideally.
A fuse with the current rating smaller than $I_{max}$ can be used for the fuse-based IRS because 58.3mA and 83.3mA flow through $P_L$ under the passive and active overcurrent attacks, respectively, as computed in Section~\ref{sec:attack}.
A fuse has to be replaced every time after it blows out.
Though replacing a fuse is simple, a circuit breaker is reusable after it isolates a VIDS during the overcurrent attack.
A system operator, however, still has to reset the circuit breaker manually.

A thermostat also disconnects two parts of a circuit if it is heated above its temperature limiting threshold and connects them again if it is cooled down below the temperature limiting threshold.
A heating coil is made of a metal composite such as nichrome that generates heat when current flows through it.
Under the passive overcurrent attack, heat is generated from the heating coil connected to CANL due to the current flowing through $P_L$.
Both heating coils generate heat under the active overcurrent attack because the current flows from $P_H$ to $P_L$.
The thermostat is heated and isolates a VIDS from the CAN bus.
Consider that the malicious code is removed from the microcontroller.
The current does not flow through the heating coil, thus cooling down the thermostat below the temperature limiting threshold.
Then, the wires from the microcontroller to the CAN bus lines are connected again without replacing any hardware components.
The heating coil and thermostat are a wire if the thermostat is closed.
Hence, the heating coil and thermostat do not thwart the voltage measurement at the microcontroller when they are attached to an analog pin as illustrated in Fig. \ref{fig:structure_heat_IRS}.

One might argue that a resettable fuse can be used instead of a thermostat because a system operator also does not need to replace the resettable fuse.
Different from fuses or thermostats that completely disconnect a wire, the resettable fuse, however, allows a leakage current in the order of a hundred mA, which is large enough to damage the microcontroller when the resettable fuse is open \cite{resettable_fuse}.
Hence, the resettable fuse should not be used in the proposed hardware-based IRSs.

\subsection{IRSs for DoS Attack, Forced Retransmission Attack, and Pulse Attack}

The output voltage from an analog pin has to be limited in order to mitigate the DoS attack, forced retransmission attack, and pulse attack since an abnormal voltage is applied to either CANH or CANL under these attacks.
Due to this voltage, current flows through the analog pins under these attacks.
Under the DoS attack, current flows from $P_L$ to the ground of the CAN transceiver when a dominant bit is transmitted.
Also, current flows from $P_H$ to the ground of the CAN transceiver under the forced retransmission attack.
Under the pulse attack, current flows through either $P_L$ or $P_H$ if the PWM signal is applied to CANL or CANH, respectively.
Hence, fuses or circuit breakers can be used to mitigate these attacks.

\begin{figure}[t!]
	\centering
	\includegraphics[width=0.2\textwidth]{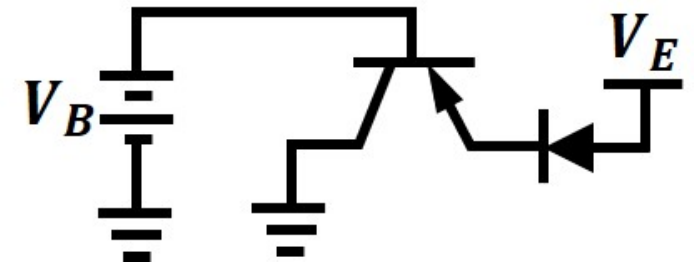}
	\caption{Test circuit for measuring current in the DoS attack.}
	\label{fig:test_circuit_DoS}
\end{figure}

In order to determine the current rating of the fuse, the current flowing through $P_L$ and $P_H$ under the DoS attack, forced retransmission attack, and pulse attack is analyzed, respectively.
Fig.~\ref{fig:test_circuit_DoS} illustrates a test circuit that emulates the voltage levels of the transistor inside the CAN transceiver under the DoS attack using a Motorola Solutions 2N2905A PNP BJT.
When the BJT is in the on-state, we measure that 281mA flows into the ground using the test circuit.
For the forced retransmission attack, we compute the current that flows through $P_H$ as $\frac{(5.0\text{V}-1.5\text{V})}{60\Omega}=$58.3mA.
Since the voltage levels under the pulse attack via CANL is the same as that under the DoS attack, the current flowing through $P_L$ becomes 281mA.
For the pulse attack via CANH, the current flowing through $P_H$ becomes 58.3mA as the forced retransmission attack.
Hence, fuses or circuit breakers with current rating less than 58.3mA mitigate these attacks.

We analyze that current flows through the analog pins under the DoS attack, forced retransmission attack, and pulse attack.
By exploiting this fact, the heating coil may heat the thermostat during these attacks, and the thermostat disconnects the extra wires.
As a result, the heat-based IRS can also mitigate these attacks by isolating a VIDS from the CAN bus.

\section{Evaluation}
\label{sec:eval}

In this section, we analyze the overcurrent attack, DoS attack, forced retransmission attack, and pulse attack on a CAN bus testbed. 
We demonstrate that the proposed hardware-based IRSs may mitigate the voltage-based attacks.

\subsection{Testbed and Equipment}

\begin{figure}[t!]
	\centering
	\includegraphics[width=0.3\textwidth]{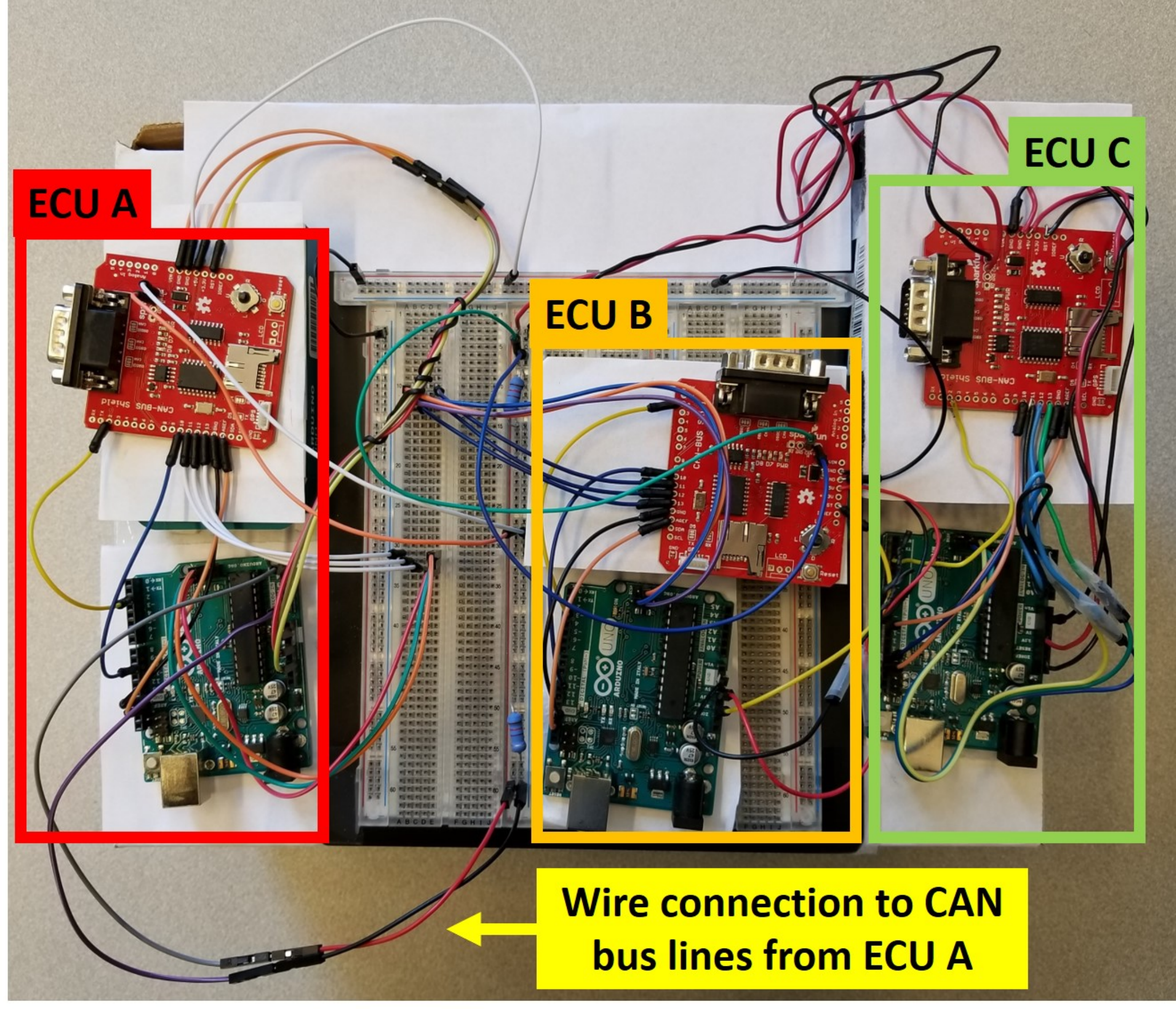}
	\caption{CAN bus testbed. The microcontroller of ECU A is connected to CANH and CANL using two extra wires. ECU A launches the proposed voltage-based attacks. ECU C transmits messages every 1s, and ECU B logs the received messages.}
	\label{fig:arduino_testbed}
\end{figure}

We implement a CAN bus testbed that consists of three ECUs as shown in Fig. \ref{fig:arduino_testbed}.
Each testbed ECU is composed of an Arduino UNO Rev3 board and a Sparkfun CAN bus shield.
The CAN bus shield uses a Microchip MCP2515 CAN controller and a Microchip MCP2551 CAN transceiver.
We set the CAN bus speed to 500kbps, which is widely used in many modern cars \cite{Shin:2016:finger}.
Since a VIDS is assumed to be installed on the microcontroller of ECU A, two analog pins A0 and A5 of the Arduino board in ECU A are connected to CANH and CANL, respectively.
ECU A is compromised and launches the proposed voltage-based attacks by changing the modes of these analog pins.
ECU C transmits messages every 1s, and ECU B logs all the received messages.

We use a Tektronix TDS2004B oscilloscope to measure voltages of the CAN bus lines and bit duration in the normal message transmission and under the voltage-based attacks.
A Keysight 34461A digital multimeter is used to measure voltage and current.
The minimum period of a PWM signal that an Arduino board can generate is 1.02ms, and the Arduino board cannot generate various voltage levels from an analog pin \cite{Arduino:UnoSpec}.
We use a Tektronix AFG3021 function generator to apply PWM signals with various periods in order to determine the minimum period for a successful pulse attack.
A Keysight U8031A power supply is used to apply various constant voltages to CANH and CANL to determine the minimum voltage levels for successful DoS attack and forced retransmission attack, respectively.

\begin{figure}[t!]
	\centering
	\includegraphics[width=0.27\textwidth]{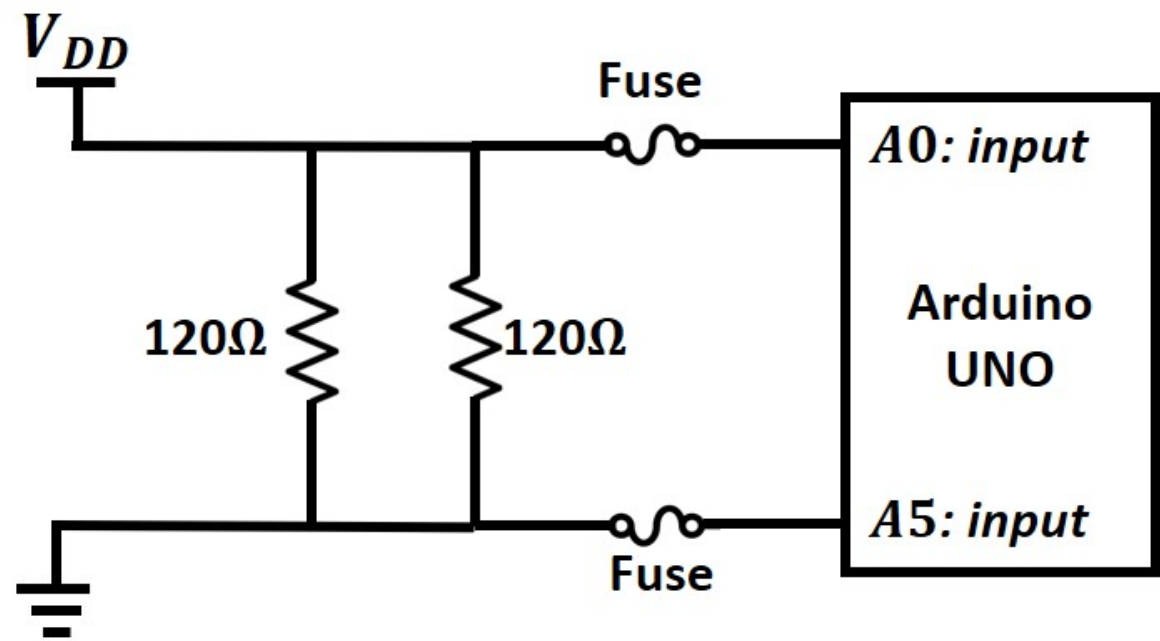}
	\caption{Test circuit for checking that the Arduino board can measure the voltage with the fuses.}
	\label{fig:test_circuit_voltage}
\end{figure}

We use Littelfuse 0326 fuses with current rating 10mA to implement the fuse-based IRS.
For the heat-based IRS, we use Sparkfun heating pad and Sensata Airpax 67L040 thermostats with temperature limiting threshold 40$^\text{o}$C.
In the test circuit as illustrated in Fig.~\ref{fig:test_circuit_voltage}, $V_{DD}$ is 5.0V, and the analog pins are in the input mode.
The microcontroller can correctly measure the voltages of CANH and CANL (i.e., 5.0V and 0.0V at A0 and A5, respectively), which indicates that the fuse does not thwart the voltage measurement at the microcontroller.
By repeating the same experiments after replacing the fuses with circuit breakers and thermostats with heating coils, respectively, we verify that our proposed IRSs do not thwart the voltage measurement.

\begin{figure}[t!]
	\centering
	\includegraphics[width=0.24\textwidth]{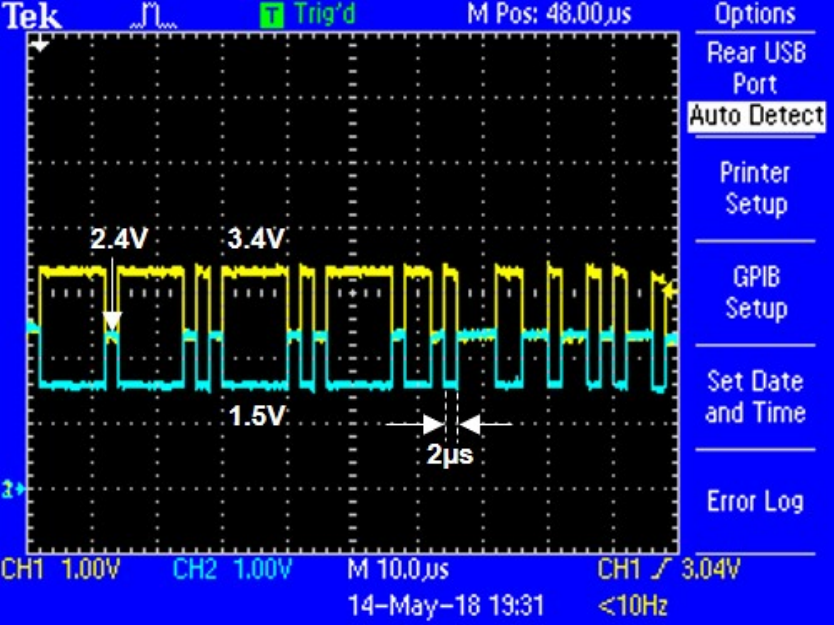}
	\caption{Voltages of CANH (in yellow) and CANL (in blue) in the normal message transmission.}
	\label{fig:normal_transmission}
\end{figure}

Fig.~\ref{fig:normal_transmission} shows the voltage levels of CANH and CANL in the normal message transmission when the message ID is 0x01 and data is 0x01.
Due to hardware characteristics of the CAN transceiver, the actual voltages of CANH and CANL are 2.4V when a recessive bit is transmitted.
When transmitting a dominant bit, the actual voltages of CANH and CANL become 3.4V and 1.5V, respectively.
The voltage levels for both dominant and recessive bits, however, meet the CAN protocol \cite{Microchip:MCP2551}.
Since the CAN bus speed is set to 500kbps, one bit is 2$\mu$s long.

\subsection{Voltage-based Attacks}

In this section, we present the experimental results of the voltage-based attacks using our CAN bus testbed.

\subsubsection{Overcurrent Attack}

\begin{figure}[t!]
	\centering
	\includegraphics[width=0.2\textwidth]{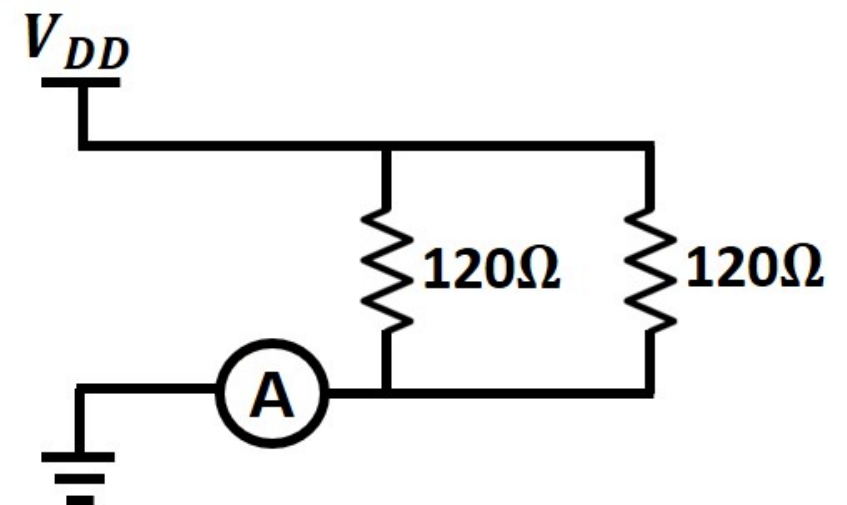}
	\caption{Test circuit to emulate the voltage levels of the CAN bus lines under the overcurrent attacks.
	$V_{DD}$ is set to 3.5V in the passive overcurrent attack and 5.0V in the active overcurrent attack, respectively.}
	\label{fig:circuit_suicide}
\end{figure}

Since we do not want to damage our testbed ECUs, we implement a test circuit as illustrated in Fig.~\ref{fig:circuit_suicide} to emulate the CAN bus under the overcurrent attack.
In the test circuit, let the ground and $V_{DD}$ represent $P_L$ of the microcontroller and the voltage of CANH, respectively.
To emulate the passive overcurrent attack, $V_{DD}$ is set to 3.5V using the power supply, and the current flowing into the ground is measured to be 58mA.
When emulating the active overcurrent attack, $V_{DD}$ is set to 5.0V, and 83mA flows into the ground.
These experimental values closely match with the theoretical values computed in Section~\ref{sec:attack}.
Since the current is supplied from the microcontroller in the active overcurrent attack, the Arduino board may not generate 83mA due to its hardware limitation.
We measure that the maximum 52mA can be provided from an analog pin of the Arduino board, which is still large enough to damage many microprocessors including Microchip ATmega328P, Renesas V850, and NXP MPC563 \cite{Arduino:UnoSpec,Renesas:V850:Datasheet,NXP:2005:MPC561}.

\begin{figure}[t!]
	\centering
	\includegraphics[width=0.35\textwidth]{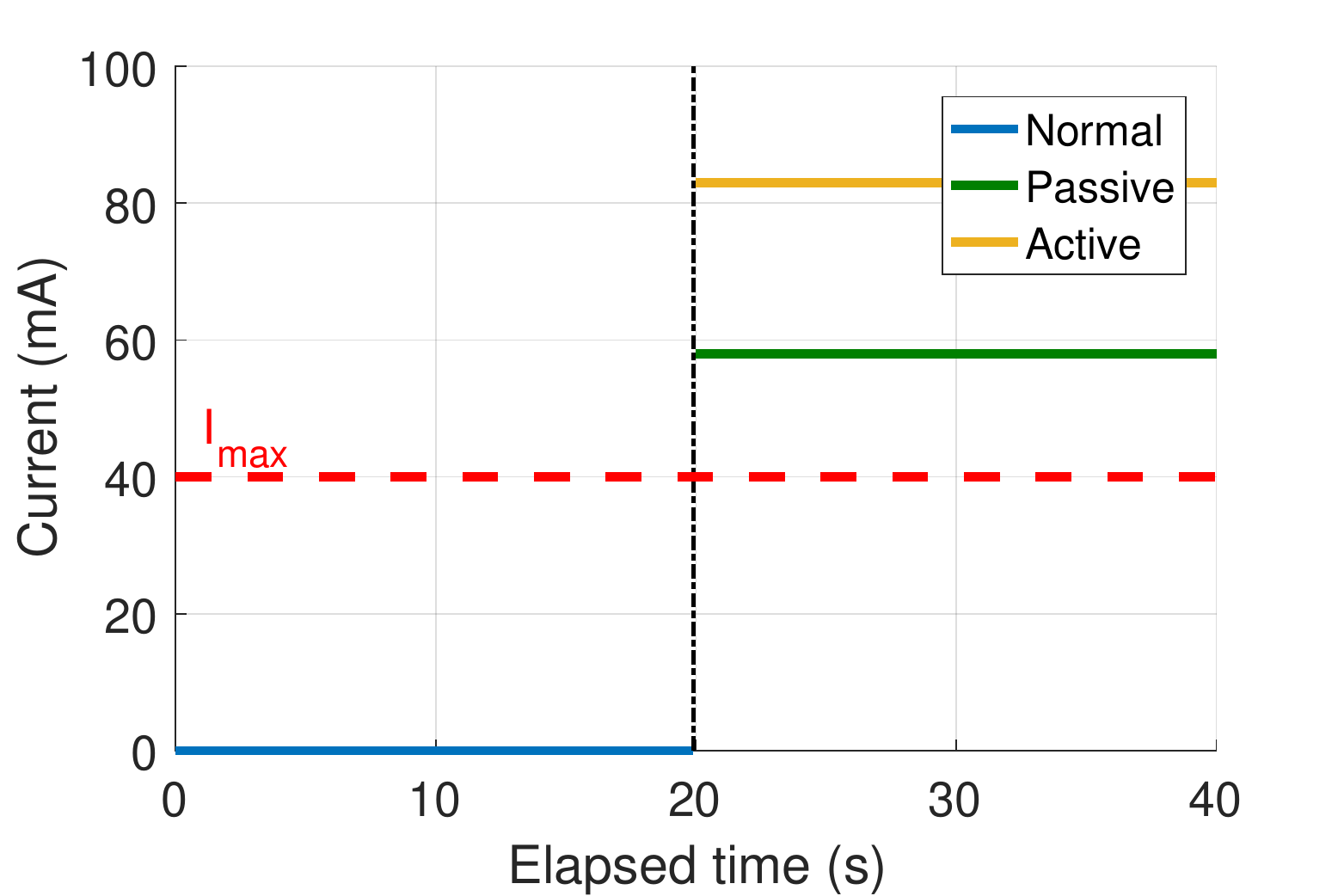}
	\caption{Current under the overcurrent attacks using the test circuit. Before launching the attacks, the current does not flow. After launching the attacks, the current greater than $I_{max}$ (i.e., 40mA) flows into the ground in both attacks, which may damage the microcontroller.}
	\label{fig:result_overcurrent_attack}
\end{figure}

Fig.~\ref{fig:result_overcurrent_attack} demonstrates the current flowing into the ground in the test circuit under the overcurrent attacks.
Both passive and active overcurrent attacks occur at 20s, which is indicated by the black dashed line.
The red dashed line represents $I_{max}$ (i.e., 40mA) of the Arduino board.
The current does not flow through the analog pins before the overcurrent attacks occur.
After the attacks, however, the current larger than $I_{max}$ flows through $P_L$, which may damage the microcontroller.

\subsubsection{DoS Attack}

\begin{figure}[t!]
	\centering
	\includegraphics[width=0.24\textwidth]{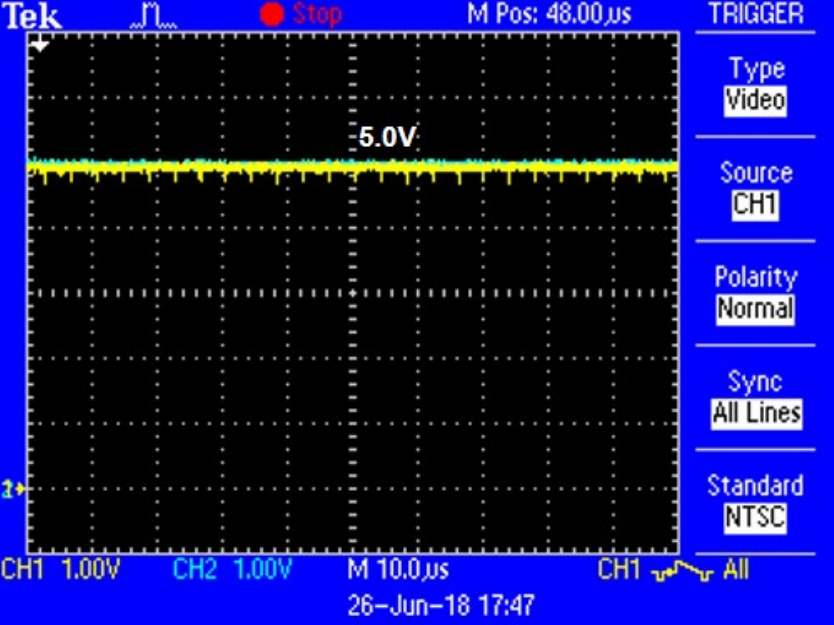}
	\caption{Voltages of CANH and CANL when the DoS attack is successfully launched by ECU A when $V_{attack,L}$ is set to 5.0V. }
	\label{fig:DoS_attack_by_Arduino}
\end{figure}

\begin{figure}[t!]
	\centering
	\includegraphics[width=0.35\textwidth]{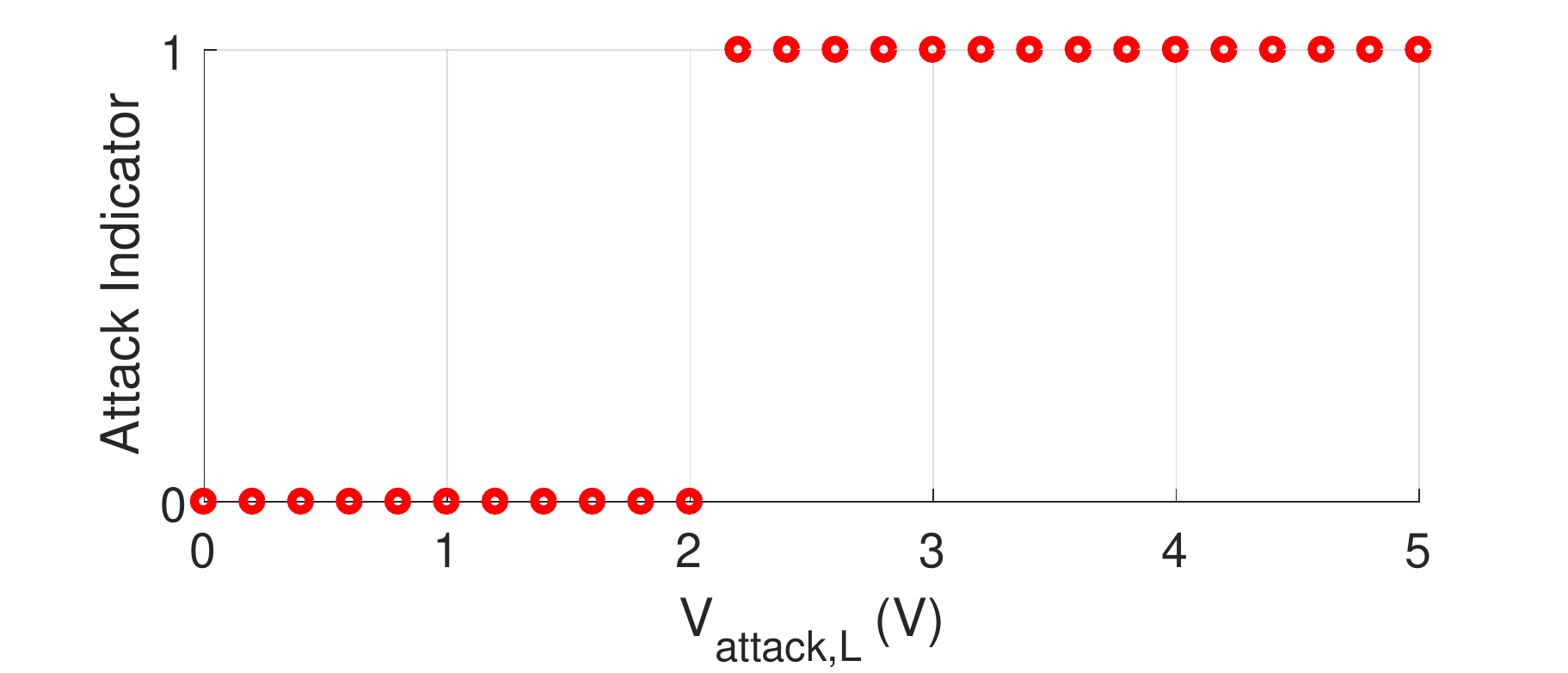}
	\caption{Minimum value of $V_{attack,L}$ that successfully launches the DoS attack. The attack indicator is either 0 if the attack fails or 1 the attack succeeds. The DoS attack becomes successful from $V_{attack,L}$=2.2V.}
	\label{fig:DoS_attack_over_voltages}
\end{figure}

Fig.~\ref{fig:DoS_attack_by_Arduino} shows that the voltages of CANH and CANL become 5.0V when $V_{attack,L}$ is set to 5.0V.
Since $V_{Diff}$ is always 0.0V, a dominant bit cannot be transmitted, which demonstrates that the DoS attack is successfully launched using ECU A.
In order to determine the minimum voltage that successfully launches the DoS attack, we connect the power supply to CANL.
$V_{attack,L}$ is increased from 0.1V to 5.0V to cover the output voltage range of many microprocessors \cite{Microchip:ATmega328PAuto,NXP:2005:MPC561}.
The attack indicator is 0 if the attack fails and 1 if the attack is successful.
As shown in Fig.~\ref{fig:DoS_attack_over_voltages}, the DoS attack is successful if $V_{attack,L}$ is between 2.2V and 5.0V since $V_{Diff}$ is always less than the decision threshold for determining a dominant bit (i.e., 0.9V for Microchip MCP2551).

\subsubsection{Forced Retransmission Attack}

\begin{figure}[t!]
	\centering
	\includegraphics[width=0.24\textwidth]{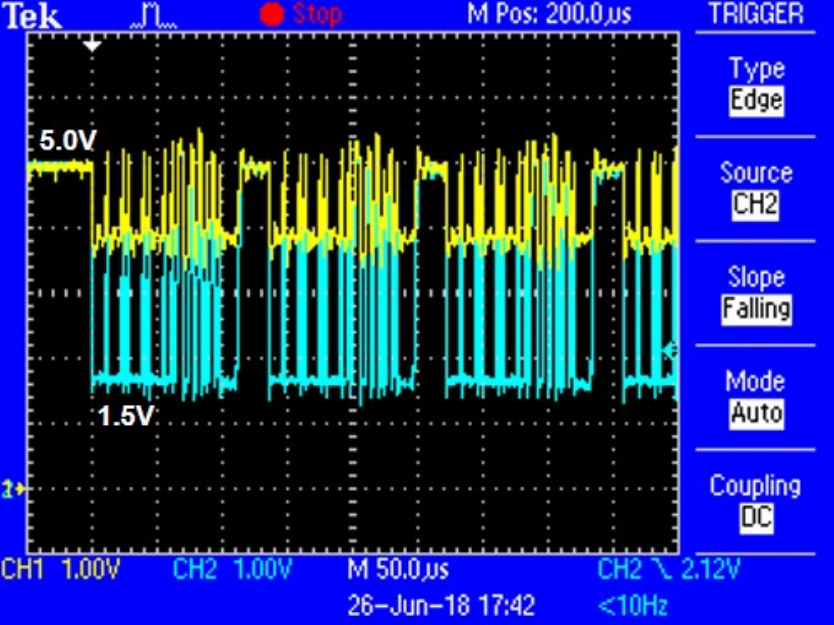}
	\caption{Voltages of CANH and CANL when $V_{attack,H}$=5.0V. The forced retransmission attack is successfully launched by ECU A. The same voltage waveform is repeated every 132$\mu$s, indicating the message retransmission.}
	\label{fig:FRA_arduino}
\end{figure}

\begin{figure}[t!]
	\centering
	\subfloat[Normal\label{fig:normal_bit_length}]{\includegraphics[width=0.24\textwidth]{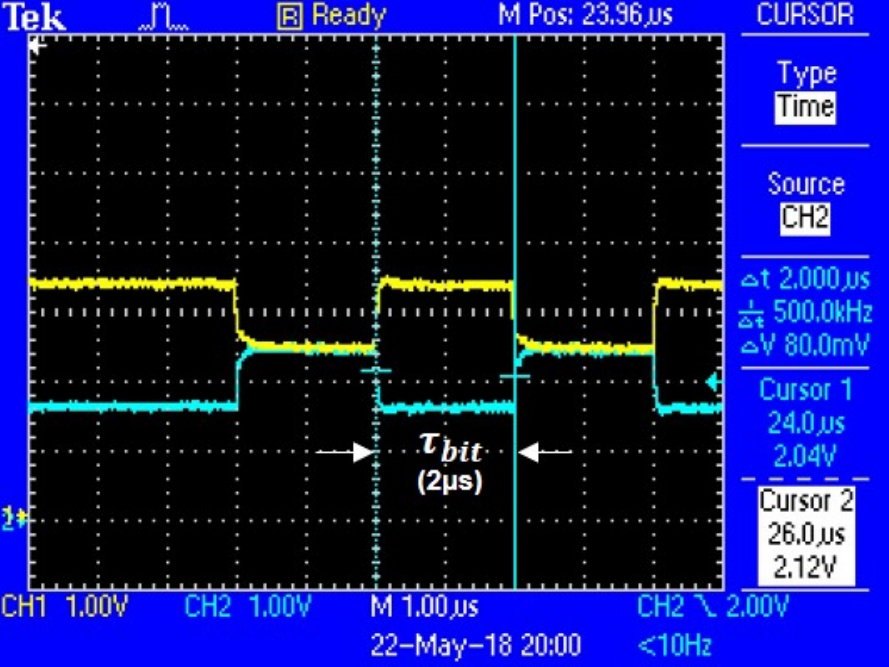}}
	\hfill
	\subfloat[$V_{attack,H}$=5.0V\label{fig:FRA_5.0_bit_length}]{\includegraphics[width=0.24\textwidth]{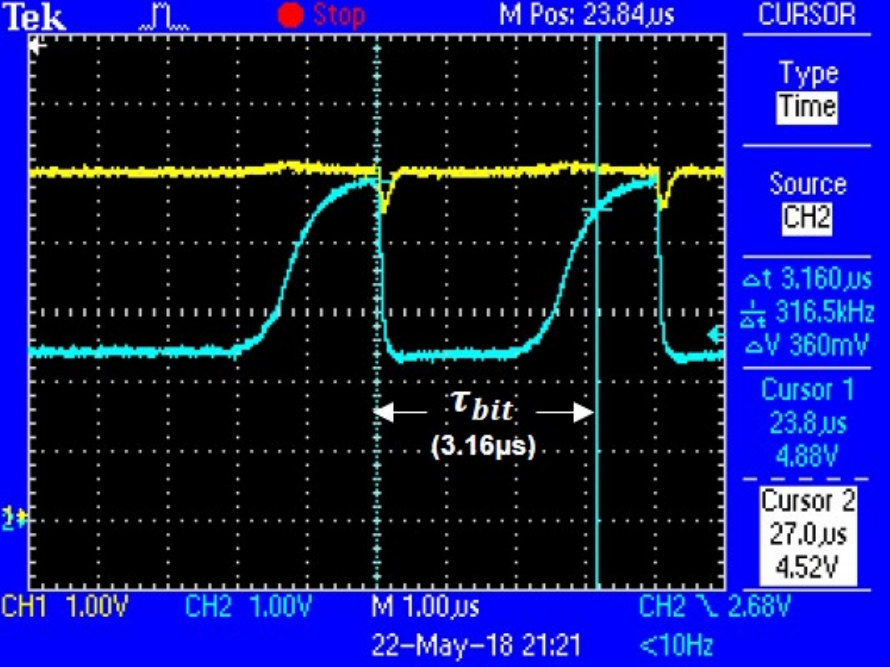}}
	\caption{$\tau_{bit}$ in the normal message transmission and under the forced retransmission attack with $V_{attack,H}$=5.0V. (a) $\tau_{bit}$ is 2$\mu$s in the normal message transmission. (b) $\tau_{bit}$ becomes 3.16$\mu$s under the attack.}
	\label{fig:FRA_bit_length}
\end{figure}

Fig.~\ref{fig:FRA_arduino} shows the voltages of the CAN bus lines under the forced retransmission attack by applying 5.0V to CANH.
Two consecutive messages are spaced by about 30$\mu$s, which indicates that the message is retransmitted because the period of the message is 1s.
The voltage of CANH could not be maintained at 5.0V since the Arduino board cannot generate current large enough to make $V_{Diff}$ greater than 3.5V.
In order to determine the minimum voltage for the successful forced retransmission attack, we increase $V_{attack,H}$ from 2.5V to 5.0V using the power supply.
The forced retransmission attack becomes successful from $V_{attack,H}$=4.5V.
Fig.~\ref{fig:FRA_bit_length} shows the bit length time $\tau_{bit}$ in the normal message transmission and under the forced retransmission attack if $V_{attack,H}$=5.0V.
While $\tau_{bit}$ is 2$\mu$s in the normal message transmission, $\tau_{bit}$ becomes 3.16$\mu$s under the forced retransmission attack.
As a result, the duration of a recessive bit violates the CAN protocol, which causes an error.
Since there are 7 transitions from a dominant bit to a recessive bit in a message with the ID 0x01 and data 0x01, Table~\ref{table:FRA_message_duration} shows the average of $\tau_{bit}$ when $V_{attack,H}$ is varied from 2.5V to 5.0V.

\begin{table}[t!]
	\footnotesize
	\centering
	\caption{Average bit length time for various values of $V_{attack,H}$ when the CAN bus speed is 500kbps.}
	\begin{tabular}{| m{1cm} | c | c | c | c | c | c |}
		\hline
		$V_{attack,H}$ & 2.5V & 3.0V & 3.5V & 4.0V & 4.5V & 5.0V\\
		\hline
		Average $\tau_{bit}$  & 2.00$\mu s$ & 2.24$\mu$s & 2.86$\mu$s & 2.98$\mu$s & 3.07$\mu$s & 3.16$\mu$s\\
		\hline
	\end{tabular}
	\label{table:FRA_message_duration}
	\normalsize
\end{table}

\subsubsection{Pulse Attack}

\begin{figure}[t!]
	\centering
	\subfloat[CANL\label{fig:pulse_attack_CANL}]{\includegraphics[width=0.24\textwidth]{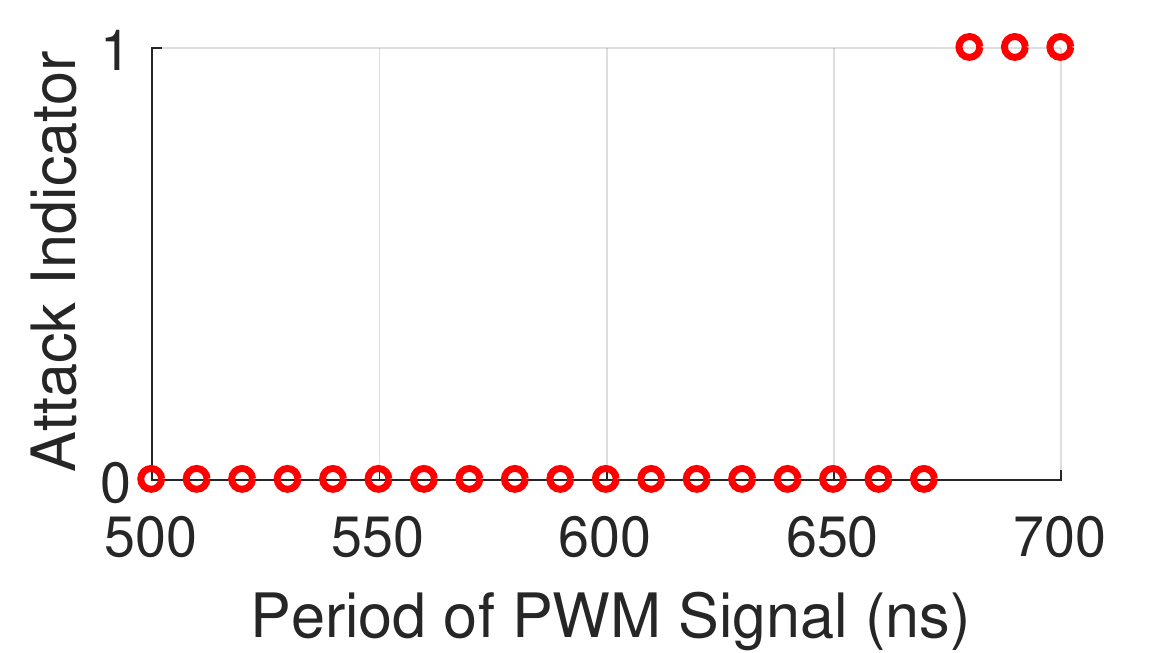}}
	\subfloat[CANH\label{fig:pulse_attack_CANH}]{\includegraphics[width=0.24\textwidth]{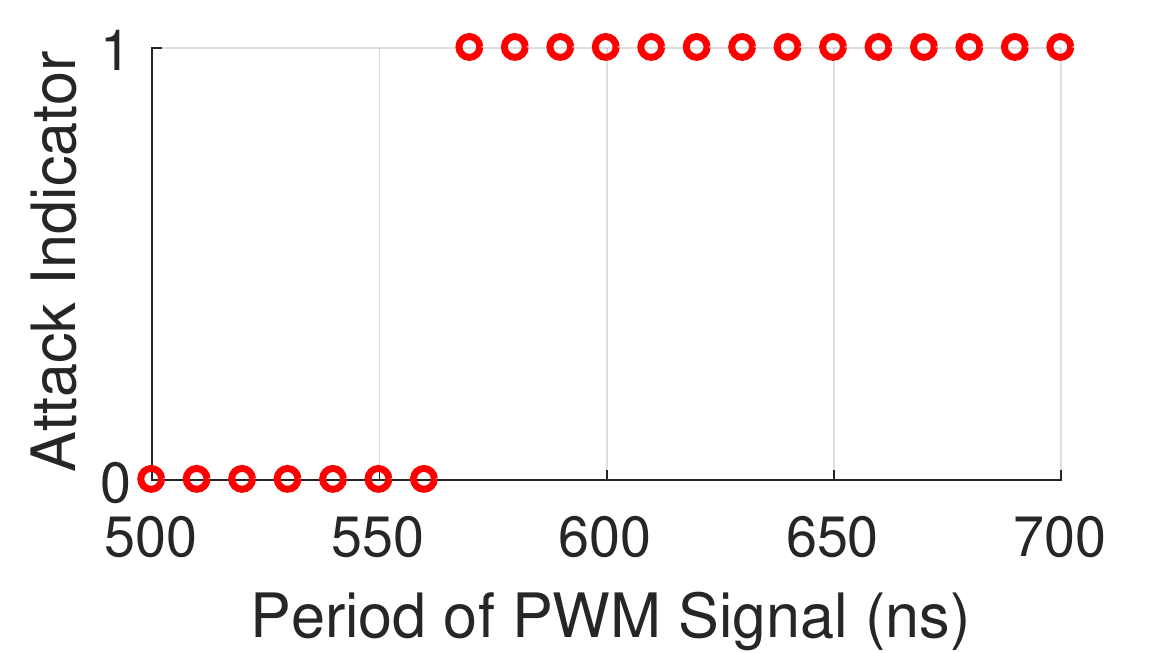}}
	\caption{Minimum period of the PWM signal that leads to a successful pulse attack. (a) The pulse attack is successful when the period is larger than 680ns if the pulse is applied to CANL. (b) The pulse attack becomes successful when the period is larger than 570ns if the pulse is applied to CANH.}
	\label{fig:pulse_attack}
\end{figure}

In order to determine the minimum period of the PWM signal that launches the pulse attack successfully, we apply the PWM signal with the duty cycle 50\%, amplitude 5.0V, and offset 2.5V (i.e., the low output voltage is 0.0V, and the high output voltage is 5.0V) to CANL using the function generator while $P_H$ is in the input mode.
We increase the period of the PWM signal from 500ns to 700ns.
Then, we repeat the same experiment after connecting the function generator to CANH where $P_L$ is set to the input mode.
Fig.~\ref{fig:pulse_attack} demonstrates that the pulse attack becomes successful if the period is longer than 680ns and 570ns when the PWM signal is applied to CANL and CANH, respectively.
The pulse attack via CANH becomes successful with a shorter period than that via CANL because the transition time increases when the voltage is applied to CANH.

\subsection{Hardware-based IRSs}

In this section, we demonstrate that the fuse-based IRS can mitigate all four voltage-based attacks.
Also, the proof-of-concept of the heat-based IRS is verified using a test circuit.

\subsubsection{Fuse-based IRS}

\begin{figure}[t!]
	\centering
	\includegraphics[width=0.37\textwidth]{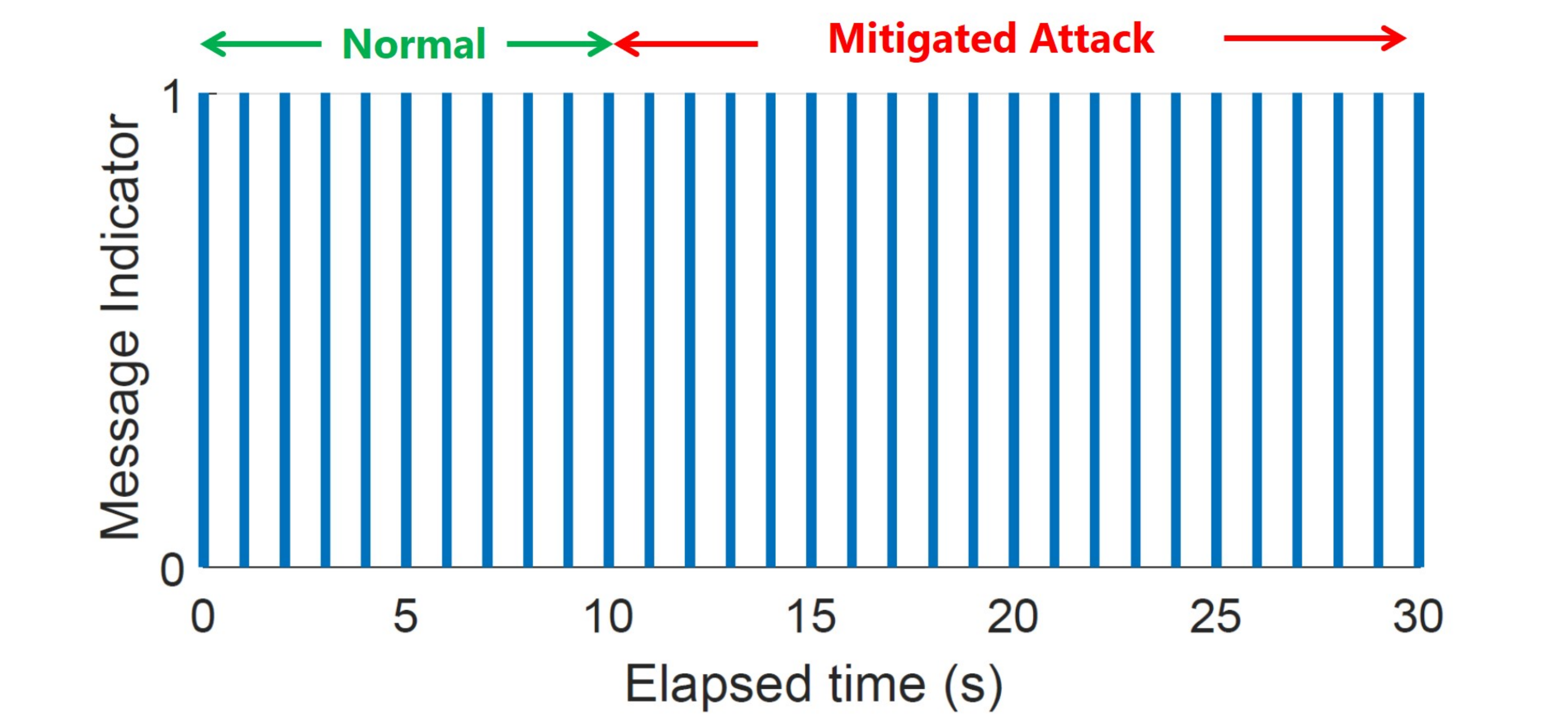}
	\caption{Message exchange through the CAN bus under the DoS attack, forced retransmission attack, and pulse attack. Each attack is launched from 10s to 30s. The fuse-based IRS mitigates all three voltage-based attacks.}
	\label{fig:voltage_attack_w_IRS_log}
\end{figure}

We implement the fuse-based IRS as illustrated in Fig.~\ref{fig:structure_IRS}.
58mA and 83mA flow through $P_L$ under the passive and active overcurrent attacks, respectively.
By using the fuse with current rating 10mA, both overcurrent attacks can be mitigated since $P_L$ is disconnected from CANL.
In order to verify that the fuse-based IRS can mitigate the DoS attack, forced retransmission attack, and pulse attack, we design an attack scenario in which ECU A launches an attack from 10s to 30s, while ECU C transmits messages every 1s.
We repeat this experiment for each of the three attacks.
For the DoS attack and forced retransmission attack, 5.0V is applied to CANL and CANH, respectively.
The period of the PWM signal is set to 100$\mu$s with the duty cycle 50\%, amplitude 5.0V, and offset 2.5V for the pulse attack.
The message indicator is 1 if the message is received at ECU B and 0 if the message transmission fails.
For all three voltage-based attacks, ECU B receives the messages every 1s as normal under each attack as shown in Fig.~\ref{fig:voltage_attack_w_IRS_log}.
Hence, the fuse-based IRS may mitigate the voltage-based attacks.

\subsubsection{Heat-based IRS}

\begin{figure}[t!]
	\centering
	\includegraphics[width=0.25\textwidth]{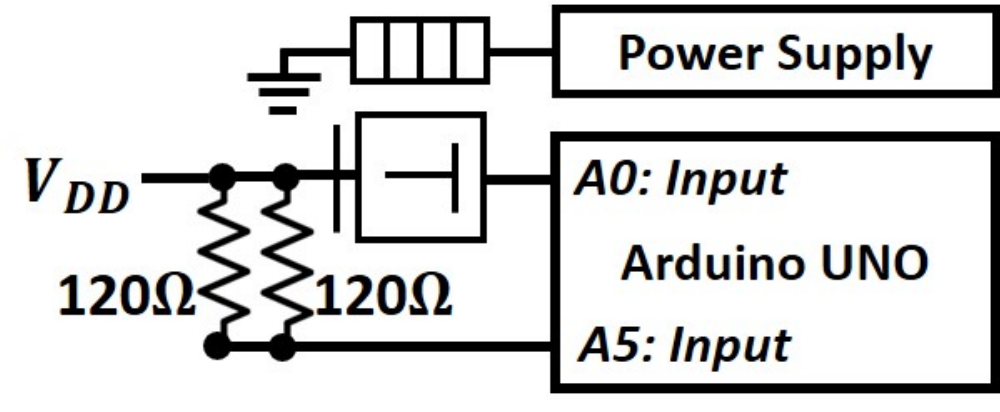}
	\caption{Test circuit to emulate the heat-based IRS. The power supply provides the current of 1A to the heating coil.}
	\label{fig:test_circuit_heat_IRS}
\end{figure}

The maximum current that the Arduino board can supply is 52mA, by which the heat may not be generated high enough to let the thermostat disconnect the wires \cite{Arduino:UnoSpec}.
We implement a test circuit that emulates operations of the heat-based IRS where the current to the heating coil is provided by the power supply instead of the Arduino board as shown in Fig.~\ref{fig:test_circuit_heat_IRS}.
Current does not flow through the heating coil when emulating the normal operation of a VIDS.
Since current flows through either $P_L$ or $P_H$ under all four voltage-based attacks, we launch an attack by letting the current of 1A flow the heating coil from the power supply under the controlled laboratory setting for the safety.
In the test circuit, A0 and A5 are in the input mode while $V_{DD}$ is set to 5.0V.
Before emulating an attack, the microcontroller measures 5.0V via A0, which means the analog pin is connected to $V_{DD}$.
After launching an emulated attack, however, the microcontroller cannot measure $V_{DD}$ at A0, which indicates that the thermostat disconnects the wire.
These experimental results demonstrate that the heat-based IRS may mitigate the voltage-based attacks.

\section{Conclusion}
\label{sec:conclusion}

In this paper, we investigated new vulnerabilities of VIDS when an adversary maliciously exploits the extra wires given to VIDS.
We proposed four voltage-based attacks that may damage the microcontroller with VIDS, block message transmission, or make a message be retransmitted by manipulating the voltage levels of the CAN bus lines.
In order to mitigate the voltage-based attacks, we developed two hardware-based IRSs that isolate VIDS from the CAN bus once the voltage-based attacks are detected. 
We demonstrated the voltage-based attacks and hardware-based IRSs using our CAN bus testbed.
Our work implies that a new attack surface might be introduced to the CAN bus by the wires that directly connect the CAN bus lines and VIDS if VIDS is compromised.
Hence, in order to provide security assurance to an automobile, defense systems based on hardware must be implemented together with VIDS if the voltage characteristics are exploited as a fingerprint of ECUs.
Although the idea of isolating a compromised VIDS using thermostats is demonstrated, implementation of the heat-based IRS in a more practical environment will be studied as future work.



\bibliographystyle{IEEEtran}
\bibliography{sagong_bib}

\end{document}